\documentclass[11pt,a4paper]{article}
\pdfoutput=1
\usepackage{graphicx}
\usepackage{jheppub}

\usepackage{amsfonts}
\usepackage{color}

\usepackage{simplewick}

\newcommand{\be}{\begin{equation}}
\newcommand{\ee}{\end{equation}}
\newcommand{\bea}{\begin{eqnarray}}
\newcommand{\eea}{\end{eqnarray}}
\newcommand{\ba}{\begin{array}}
\newcommand{\ea}{\end{array}}

\def \nn {\nonumber}
\newcommand{\eq}[1]{(\ref{#1})}

\newcommand{\Tr}{\mbox{Tr}}

\newcommand{\cA}{{\cal{A}}}

\newcommand{\cF}{{\cal{F}}}
\newcommand{\cG}{{\cal{G}}}

\newcommand{\cO}{{\cal{O}}}

\newcommand{\beq}{\begin{eqnarray}}
\newcommand{\eeq}{\end{eqnarray}}
\newcommand{\bes}{\begin{subequations}}
\newcommand{\ees}{\end{subequations}}

\title{Decoherence of Topological Qubit in Linear and Circular Motions: Decoherence Impedance, Anti-Unruh and Information Backflow}

\author[a]{Pei-Hua Liu}
\emailAdd{lphhpl@gmail.com}
\affiliation[a]{Department of Physics, National Taiwan Normal University, Taipei, 116, Taiwan}
 \author[a]{and Feng-Li Lin} \emailAdd{linfengli@phy.ntnu.edu.tw} 
  
\abstract{In this paper, we consider the decoherence patterns of a topological qubit made of two Majorana zero modes in the generic linear and circular motions in the Minkowski spacetime. We show that the reduced dynamics is exact without Markov approximation. Our results imply that the acceleration will cause thermalization as expected by Unruh effect. However, for the short-time scale, we find the rate of decoherence is anti-correlated with the acceleration, as kind of decoherence impedance. This is in fact related to the ``anti-Unruh" phenomenon previously found by studying the transition probability of Unruh-DeWitt detector. We also obtain the information backflow by some time modulations of coupling constant or acceleration, which is a characteristic of the underlying non-Markovian reduced dynamics. Moreover, by exploiting the nonlocal nature of the topological qubit, we find that some incoherent accelerations of the constituent Majorana zero modes can preserve the coherence instead of thermalizing it.}  

\begin{document}

\maketitle

\section{Introduction}

     The Unruh-DeWitt (UDW) model \cite{Unruh:1976db,UDW} and many of its variations, see for example \cite{Crispino:2007eb} and reference therein, have been used to explore the novel thermal effect, i.e., Unruh effect \cite{Unruh:1976db} felt by a constantly accelerating qubit/particle coupled to the environmental field. This phenomenon shed light on our understanding about Hawking radiation \cite{Hawking:1974sw} and lately inspired the related study on relativistic quantum information \cite{Peres:2002wx}, for examples see \cite{Alsing & Milburn,FuentesSchuller:2004xp,Alsing:2006cj,Lin:2008jj,MartinMartinez:2011mw,Ostapchuk:2011ud,Richter:2015wha}. 

    However, most of studies on the UDW model are based on the time-dependent perturbation theory in order to calculate the transition rate of the UDW detector, which shows a thermal pattern as expected by the Unruh effect. In fact, the transition rate can also be obtained beyond the scheme of perturbation theory by calculating the reduced dynamics of the UDW detector, moreover from which the decoherence pattern of the UDW detector can be obtained. The full reduced dynamics if obtained can reveal the non-Markovian behavior which is ignored in the perturbation scheme \cite{Lin & Hu}.  The difficulty for such consideration lies on solving the exact master equation for the reduced dynamics of the UDW detector. If a general scheme of exact dynamics exists, one can even consider the decoherence patterns of the UDW detector under the arbitrary motion. Note that in the perturbation scheme the Unruh-like effect for the circular motion has been considered in \cite{Bell:1982qr,Bell:1986ir,circularUnruh}, for the oscillatory motion in \cite{Doukas:2013noa} (from exact reduced dynamics), non-uniform acceleration in \cite{Obadia:2007qf,Kothawala:2009aj,Barbado:2012fy},  and for finite duration acceleration in \cite{Svaiter:1992xt,Brenna:2015fga}.  Especially, in \cite{Brenna:2015fga} a generic ``anti-Unruh" phenomenon is observed, i.e., for a UDW detector coupled to the environment field only in a finite duration (but long enough to satisfy the KMS condition), its KMS temperature decreases with acceleration in certain regimes. 
    
    To bypass the difficulty of solving the exact master equation and obtain the exact reduced dynamics of a qubit in arbitrary motion, in this paper we consider the topological qubit made of pairs of moving constituent Majorana zero modes.  These Majorana zero modes  can be realized as the edge modes of the 1D topological superconductor such as the Kitaev's fermion chain \cite{Kitaev,Alicea,Fisher}.  These edge modes are robust against fermion-parity-preserving perturbations, in a similar way as for the most topological insulators/superconductors \cite{HasanKane,Qi2011,SPT-W}. However, the robustness will not be preserved if the Majorana zero modes couple to the environment via interactions violating the fermion-parity symmetry. In such a case, the quantum information of these zero modes will escape into the environment. This then causes quantum decoherence of a topological qubit made of two spatially-separated Majorana zero modes.

\begin{figure}
\includegraphics[width=.5\columnwidth]{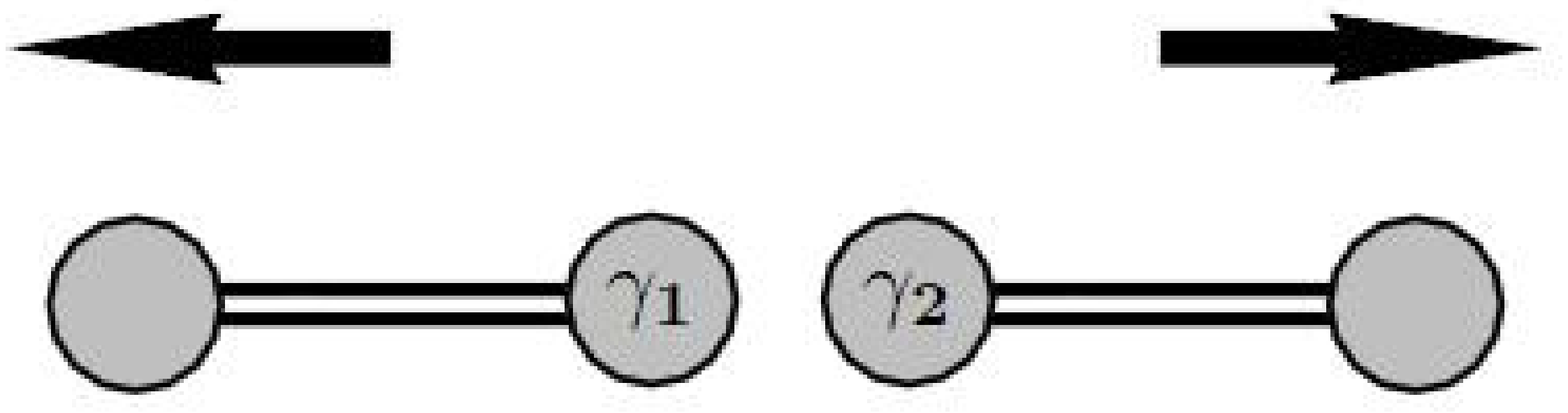}
\includegraphics[width=.5\columnwidth]{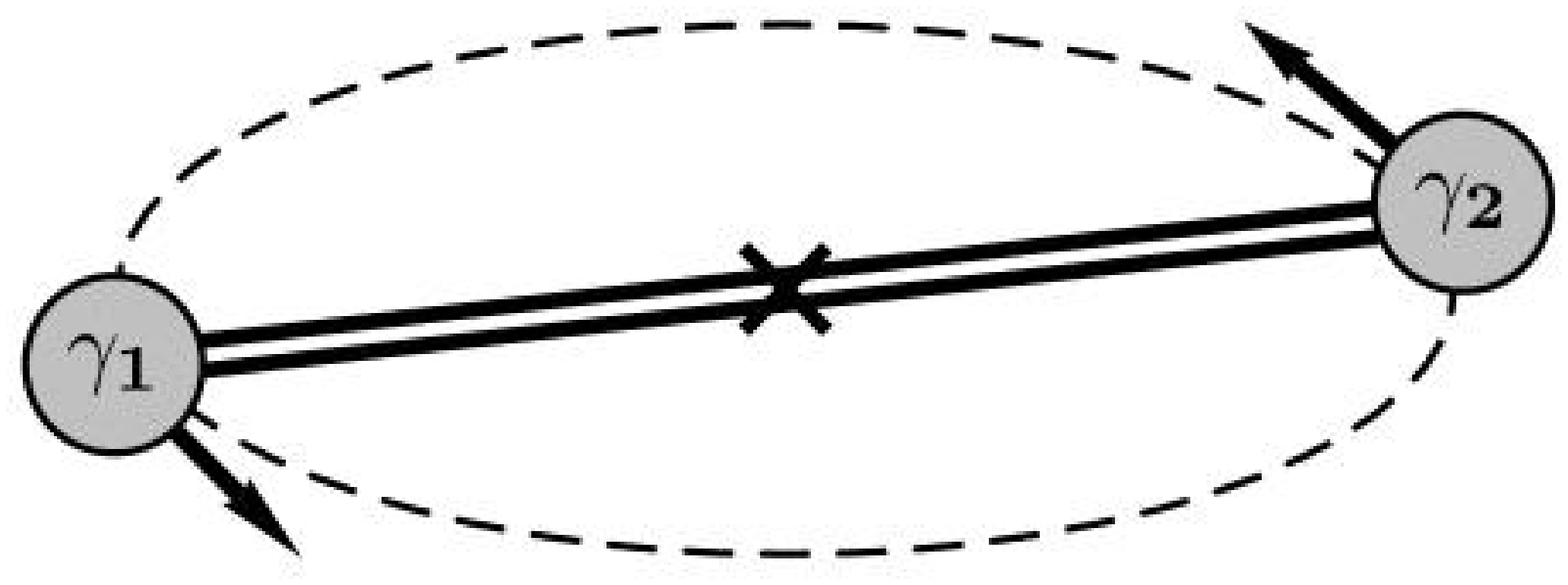}
\caption{Left: A pair of topological qubits with their constituent Majorana zero modes in incoherent  linear boost or acceleration.   Right: A topological qubit with its constituent Majorana zero modes in a coherent circular motions. In both cases, the Majorana zero modes $\gamma_{1,2}$ form a topological qubit of which we study the effect of the motions to the decoherence pattern.}
\label{cartoon}
\end{figure}

    The exact reduced dynamics for such (static) topological qubits has been obtained without solving the master equation in \cite{MajDeco,Chamon}. The solvability of the reduced dynamics for the topological qubit is due to the spatial separation of the constituent Majorana zero modes so that the influence functional is further restricted by the additional locality constraint which is absent for the usual qubit. Based on this solvability, in this paper we will develop a formalism by generalizing the one in \cite{Brown:2012pw} for the usual UDW detector to obtain the reduced dynamics for the topological qubit, the constituent Majorana zero modes of which are in arbitrary motions. Especially,  we will study the decoherence pattern of the topological qubit with its constituent Majorana zero modes moving as shown in Fig. \ref{cartoon}.  
    
    In the left panel  of  Fig. \ref{cartoon}, two wires hosting the edge Majorana zero modes are either in boost or in acceleration. The pair of the specified Majorana zero modes $\gamma_1$ and $\gamma_2$ move incoherently (i.e., with different accelerations) but form a topological qubit.  On the other hand, in the right panel, two Majorana zero modes on the edge of a wire move coherently (i.e., with the same acceleration) in a circular motion and again form a topological qubit.  We will then study the effect of these motions to the decoherence patterns of the topological qubit.

    Despite that the topological qubit is novel and non-local, the effect of the (acceleration) motions to its decoherence patterns may bear some generic features shared by the usual qubit in the similar motions. In some sense, it can be thought as the manifestation of the Unruh effect in the decoherence.  Here we list what we found in this work, which we think should be generic:
    
\begin{itemize}
 
 \item Thermalization: We find that the acceleration does cause thermalization as expected by the Unruh effect. This can be seen by the complete decoherence of the accelerating topological qubit in the super-Ohmic environment as this kind of qubits without acceleration are robust against decoherence. However, the influence functional is different from the thermal one. It is worthy of further study to see if the influence functional of the usual qubit in the constant acceleration is the same as the thermal one or not. 
 
 \item Anti-Unruh: In our formalism we can obtain the full transition probability of the topological qubit as the UDW detector, which can be used to characterize Unruh effect. Follow the scheme in \cite{Brenna:2015fga} by decoupling the UDW detector from the environmental field after finite time duration, we can find from the transition probability the aforementioned ``anti-Unruh" phenomenon for the accelerating topological qubit in both linear and circular motions. Our result is exact without Markov approximation as done in \cite{Brenna:2015fga} and thus justifies the ``anti-Unruh" phenomenon is generic.  
 
 \item Decoherence impedance: By studying the decoherence patterns we find some novel feature. it seems that the topological qubit is resisting the enforced decoherence due to the sudden change of the motion status such as acceleration or boost. We called this tendency the {\it decoherence impedance}, namely, the initial rate of decoherence is anti-correlated with the size of change, such as acceleration. In the late time, this impedance effect will be taken over by thermalization so that in the mean time the overtaking of the decoherence rates happens. 
 
    In fact, both the decoherence patterns and the transition probability are all dictated by the influence functional. We would expect the decoherence impedance and ``anti-Unruh" phenomenon imply each other as both are the short-time non-equilibrium effects. This is indeed what we observe in this work.  
  
  \item Information backflow and time modulation: As our reduced dynamics is exact without Markov approximation, we shall expect some non-Markovian behaviors characterized by the information flow. We thus apply the time modulations to the coupling constant (called ``switching function" in this paper) and the acceleration to see if some modulations can invoke information backflow.  We find that not all modulations will bring out information backflow but some does and sometimes even ensures the robustness against decoherence.

\end{itemize}

   Moreover, by really exploiting the non-local feature of the topological qubit, in this paper we will  consider the frame dependence issues when both Majorana zero modes move incoherently.  We indeed find some novel feature: some incoherent accelerations will instead preserve the coherence of the topological qubit so that it will not be thermalized by the accelerations. This is a phenomenon peculiar for the topological qubit as its constituents can move differently.

    The exact solvability of the reduced dynamics even with the nontrivial motions of the topological qubit and the results presented in this paper demonstrate the power of marrying topological order with relativistic quantum information. It is interesting to see if some of the ``generic" features aforementioned will also occur for the usual qubit. The answer to this question will definitely shed new light on our understanding of relativistic quantum information.

    This paper is organized as follows: in the next section we will briefly review the formulation of  the topological qubit and its the reduced dynamics developed in \cite{MajDeco}. In the section \ref{section III} we develop the formalism to study the reduced dynamics for a moving topological qubit by generalizing the one given in \cite{Brown:2012pw} for the usual UDW detector. Especially, in \ref{section IIIC} we obtain the exact transition probability of the topological qubit beyond the Markovian approximation. Based on our developed formalism, we will present the decoherence patterns for various linear motions  in section \ref{sec IV}, and for various circular motions  in section \ref{sec V}.  We then conclude our paper in section \ref{secCon}.  Besides, four technical appendices are given: Appendix \ref{app A} is  for the evaluation of the ``influence functional" for the reduced dynamics of the topological qubit, and Appendix \ref{app B} is about defining Schwinger-Keldysh Green functions and the related spectral density used in this paper. In Appendix \ref{app C} the transition probability in the Markov approximation is derived. In Appendix \ref{app D} we consider the causal condition in the comoving coordinates so that the time directions of the two incoherent Majorana zero modes are ensured to be the same.

\section{Topological Qubits and their reduced dynamics}
 
   In \cite{MajDeco} the topological qubits made of pairs of Majorana modes have been introduced and their decoherence dynamics were then studied. A novel feature for these peculiar qubits is their robustness against quantum decoherence in the super-Ohmic environments. In this section we will briefly review the setup and the resultant reduced dynamics, whose formal form applies also to the cases of moving Majorana modes as will be shown in the next section.

    A topological qubit is made of a pair of Majorana modes which can be spatially separated so that each Majorana mode interact independently with local environment, i.e., the interaction Hamiltonian
\be\label{intV}
V:= \gamma_1 O_1 + \gamma_2 O_2
\ee 
where $\gamma_{1,2}$ are the Majorana operators, i.e., they satisfy the Clifford algebra: $\gamma_1^2=\gamma_2^2=1$ and $\gamma_1\gamma_2+\gamma_2\gamma_1=0$.  These Majorana modes can be realized as the gapless edge modes of topological superconductors such as the Kitaev's chain \cite{Kitaev,Alicea,Fisher}. Thus, these Majorana modes are degenerate with zero kinetic energy, i.e., zero modes. Note $O^{\dagger}_i=-O_i$ so that $V^{\dagger}=V$.

   Furthermore, in the interaction picture we further assume that the evolving operators $\cO_i(t):=e^{i H_E t} O_i e^{-i H_E t}$ with $H_E$ the environmental Hamiltonian obey the locality constraint 
\be\label{topo-c}
\langle \cO_1(t) \cO_2(t') \rangle =0
\ee
where the vev is taken with respect to the environmental ground state denoted by $\rho_{E,0}$, i.e., $\langle \cO_i(t) \cO_j(t') \rangle:=\Tr_E \rho_{E,0} \cO_i(t) \cO_j(t')$. 

   The locality constraint \eq{topo-c} is equivalent to the usual one in the context of quantum field theory, which says that two observables are commuting if they are spacelike separated.  We will assume that the interaction $H_E$ is local and the two operators appearing in \eq{topo-c} are spacelike separated in our setup of topological qubits  so that \eq{topo-c} is satisfied.

   We can pack together a pair of Majorana zero modes to form a fermionic qubit though they are physically separated, i.e., 
\be
\gamma_1:=a^{\dagger}+a\;, \qquad \gamma_2:=i\;(a^{\dagger}-a)
\ee
so that $a^{\dagger}$ and $a$ are respectively the creation and annihilation operators  of a fermionic 2-level state, and satisfy $a^{\dagger}a+aa^{\dagger}=1$. 

   Due to the relation \eq{topo-c}, the reduced dynamics of the topological qubit is far easier to be solved than the usual fermionic qubit. The latter has been solved in \cite{Zhang} via the coherent state representation of path integral but the results cannot be put into a closed form. On the other hand, given the initial state of the topological qubit
 \beq \label{frho-ini}
\rho_M(t=0)=\left(\begin{array}{cc} \rho_{00} \;
& \; \rho_{01} \\ \rho_{01}^* \; &  1-\rho_{00} \end{array}\right)\;,
\eeq 
its reduced dynamics has been solved in \cite{MajDeco} with the simple closed form
\beq \label{frho-2}
\rho_M(t)=\left(\begin{array}{cc}{1\over 2}+(\rho_{00}-{1\over 2})e^{\mathcal{I}_1(t)+\mathcal{I}_2(t)} &e^{\mathcal{I}_2(t)} \textrm{Re} \; \rho_{01}+i e^{\mathcal{I}_1(t)} \textrm{Im}\; \rho_{01} \\  e^{\mathcal{I}_2(t)} \textrm{Re}\; \rho_{01}-i e^{\mathcal{I}_1(t)} \textrm{Im}\; \rho_{01} & {1\over 2}-(\rho_{00}-{1\over 2})e^{\mathcal{I}_1(t)+\mathcal{I}_2(t)} \end{array}\right)  
\eeq 
where the ``influence functional" associated with the $i$-th Majorana zero mode is 
\be\label{influence-alpha}
{\mathcal I}_i(t):= 2  \int^t d\tau \int^t d\tau' \; \overline{G}_{i,sym}(\tau-\tau')
\ee
with 
\be\label{dGsym}
\overline{G}_{i,sym}(t-t')={1\over 2}\big( \langle \cO_i(t) \cO_i(t') \rangle + \langle \cO_i(t') \cO_i(t) \rangle \big)\;.
\ee
Here, $\overline{G}_{i,sym}$ is the so-called Majorana-dressed symmetric Green function, which is different from the usual fermionic one, i.e., $G_{i,sym}(t-t')={1\over 2}\left( \langle \cO_i(t) \cO_i(t') \rangle - \langle \cO_i(t') \cO_i(t) \rangle \right)$. The sign change is due to the Majorana dressing so that time-ordering for $\cO_i$ changes from the fermionic one to the bosonic one. 

   The derivation of \eq{frho-2} and \eq{influence-alpha} has been sketched in \cite{MajDeco}, which involves the evaluation of the cosh/sinh correlators $\langle \cosh \cO_i(t) \cosh \cO_i(t) \rangle$ and $\langle \sinh \cO_i(t) \sinh \cO_i(t) \rangle$.  However, a not so rigorous procedure of re-exponentiation of $\langle \cO_i(t) \cO_i(t) \rangle$ is invoked to obtain the cosh/sinh correlators.   In appendix A, we re-derive the result by invoking the merging formula of operator product expansion  given in \cite{Polchinski}.
   
    Based on \eq{frho-2} and \eq{influence-alpha}, in \cite{MajDeco} we showed that reduced density matrix \eq{frho-2} relaxed to Gibbs state, i.e., $e^{\mathcal{I}_i(t\rightarrow \infty)} \longrightarrow 0$ if the spectral density for the excitations created by operator $\cO_i$ is sub-Ohmic and Ohmic, i.e., the Fourier transform of the symmetric Green function, $\cA_i(\omega)\sim \omega^Q$ for $Q\le 1$.  However, for the super-Ohmic spectral density, i.e., $Q>1$ the topological qubit is robust against decoherence, i.e., $e^{\mathcal{I}_i(t\rightarrow \infty)}\ne 0$.

\section{Formalism for the reduced dynamics of a moving topological qubit}\label{section III}

     As mentioned in the Introduction, most of the studies for the Unruh effect in the context of relativistic quantum information are about the entanglement degradation due to the constant acceleration of a pair of qubits of particles. In such cases, the reduced dynamics and the entanglement entropy are obtained by performing Bogoliubov transformation to relate the Minkowski vacuum and the Rindler one. This method, however, cannot be generalized to arbitrary motions. On the other hand, for most of the studies on UDW detector perturbation scheme is usually adopted, which is the Markovian approximation. Therefore, the interesting non-Markovian behaviors of the reduced dynamics such as information backflow will be missed. To retain the non-Markovinity one needs the non-perturbative treatments as the one developed in \cite{Hu:1991di,Zhang,ZhangPRL,MajDeco}.

 In this section we develop a formalism to deal with the reduced dynamics of a moving topological qubit.  Basically, we are applying the formalism for UDW detector given in \cite{Brown:2012pw} to the topological qubit and its reduced dynamics developed in \cite{MajDeco}.   The key idea for the formalism of \cite{Brown:2012pw} is to take care of the motion effect by performing proper coordinate transformations on the evolving Hamiltonians for either the UDW detector or the environmental fields. Therefore, it can be applied to the UDW detector in arbitrary motion. Of course, the exact solvability of the reduced dynamics will depend on the form of total Hamiltonian and the related dynamical constraints. For the topological qubit, as seen below the additional locality constraint \eq{topo-c} on the environmental correlator will make the reduced dynamics exactly solvable even when the constituent Majorana zero modes are in motion.

\subsection{Formalism} \label{formalism-time}
 
   Let us denote the environmental Hamiltonian by $H_E$, which we assume to be constant in time and space. On the other hand, the environmental operators to which the Majorana modes couple are local operators so that they are functions of space coordinate only, i.e., $O[\vec{x}]$. For simplicity, we will only consider the Majorana zero modes so that they have zero kinetic Hamiltonian for them, i.e., $H_M=0$. Therefore, the total Hamiltonian in the Schrodinger picture is 
\be
H^S_T=H_E+ V
\ee
where $V$ is given by \eq{intV}.
\

   Consider that the Majorana zero modes are in motion, and assume their trajectories are parameterized by their own proper time $\tau_i$, $i=1,2$, i.e.,
\be\label{proper-frame}
t=t(\tau_i)\;, \qquad \vec{x}=\vec{x}(\tau_i)\;.
\ee
To further evolve the whole system, we should choose a local observer and rewrite the Hamiltonian according to the observer's proper time by using the relations \eq{proper-frame}.

If we choose the local observer which is static with respect to the environment, i.e., $t$ is the proper time, then in this frame the total Hamiltonian {\it  in the interaction picture} is
\be
H^{D,t}_T=H_E+V^t
\ee
with
\be
V^t:= \sum_{i=1,2} {d\tau_i(t) \over dt} \lambda_i[\tau_i(t)]\; \gamma_i \; O_i[t,\vec{x}(\tau_i(t))] :=\sum_i \gamma_i \; \cO^{t}_i(t)   \label{OI-1}
\ee 
where we also introduce the switching functions $\lambda_{1,2}$ depending on the local proper times $\tau_{1,2}$, repsectively.   Moreover, in the above we have introduced the operators in the interaction picture, i.e.,
\be
O_i[t,\vec{x}]:=e^{i H_E t}\;  O_i[\vec{x}] \; e^{-i H_E t} \;.
\ee

  On the other hand, if we choose the local observer comoving with the $k$-th Majorana zero mode, then the total Hamiltonian {\it  in the interaction picture} is
\be
H^{D,\tau_k}_T=H_E^{\tau_k}+V^{\tau_k}
\ee
with
\be
H^{\tau_k}:={d t(\tau_k) \over d\tau_k} H_E
\ee
and
\bea
V^{\tau_k}&:=&\lambda_k[\tau_k] \; \gamma_k\; O_k[t(\tau_k),\vec{x}(\tau_k)] + \sum_{i \ne k}{d\tau_i(t(\tau_k)) \over d\tau_k} \lambda_i[\tau_i(t(\tau_k))]\; \gamma_i\; O_i[t(\tau_k),\vec{x}(\tau_i(t(\tau_k)))] \nn  \\
&:=&\sum_{i} \gamma_i \; \cO^{\tau_k}_i(\tau_k)\;. \label{OI-2}
\eea
Here, similarly $O_i[t(\tau_k),\vec{x}]=e^{i H^{\tau_k}_E \tau_k}\;  O_i[\vec{x}] \; e^{-i H^{\tau_k}_E \tau_k}$.

    To be concise, we can unify the above once a local frame is chosen. Denote the local proper time denoted by $\tau$, i.e., $\tau$ can be $t$, or one of $\tau_i$, and the total Hamiltonian  in the interaction picture is written as 
\be
H^{D,\tau}:=H_E^{\tau} + V^{\tau}
\ee
with 
\be
V^{\tau}:=\sum_i \gamma_i \; \cO^{\tau}_i(\tau)
\ee
where $O_i^{\tau}$ are defined in either \eq{OI-1} or \eq{OI-2}. 
 
 Then,  the density matrix (in the interaction picture) evolves as 
\be
\rho^D(\tau)=U(\tau) \; \rho_0 \; U^{\dagger}(\tau)
\ee
where we will assume the initial state $\rho_0$ at $\tau=0$ in the form of direct product of the states of the topological qubit and the environment, i.e., $\rho_{M,0} \otimes \rho_{E,0}$ where the environmental state is assumed to be either vacuum or thermal state of non-zero temperature. The evolution operator is formally given by
\be
U(\tau)=\textrm{T} \; e^{-i\int^{\tau} \; V^{\tau}(\tau') d\tau'}
\ee
where $\textrm{T}$ denotes the time-ordering for fermionic fields. 

We can then obtain the reduced density matrix for the topological qubit by tracing over the environmental states, i.e.,
\be\label{rhoDM}
\rho_M(\tau)=\Tr_E \; e^{-i H_E^{\tau} \tau} \; \rho^D(\tau) \; e^{i H_E^{\tau} \tau} =\Tr_E \; \rho^D(\tau)\;.
\ee
To further simplify \eq{rhoDM} we should reduce $U$ by using the Clifford algebra of $\gamma_i$'s. We then arrive
\be
U(\tau)=\textrm{T} \; e^{-i\sum_{1=1,2} \gamma_i {\bf O}_i(\tau)}=\mathcal{T} \; \Pi_{i=1,2} \big(\cosh {\bf O}_i(\tau)-i \gamma_i \sinh {\bf O}_i(\tau) \big)
\ee
where 
\be
{\bf O}_i(\tau)=\int^{\tau} d\tau' \; \cO_i^{\tau}(\tau')
\ee
and $\mathcal{T}$ denotes the time-ordering for bosonic fields. The change of time-ordering from $\textrm{T}$ to $\mathcal{T}$ is due to the removal of the dressing of $\gamma_i$ from ${\bf O}_i$, e.g., 
\bea
\textrm{T} \; \gamma_1 {\bf O}_1(\tau) \gamma_1 {\bf O}_1(\tau)&=&-\int_0^{\tau}d\tau_1 \int^{\tau}_0 d\tau_2 [ \Theta(\tau_1-\tau_2)\cO^{\tau}_1(\tau_1) \cO^{\tau}_1(\tau_2) +(\tau_1 \leftrightarrow \tau_2) ]\nn\\
&=& -\int_0^{\tau}d\tau_1 \int^{\tau}_0 d\tau_2 \; \mathcal{T} \; \cO^{\tau}_1(\tau_1) \cO^{\tau}_1(\tau_2):=- \mathcal{T} \; {\bf O}_1(\tau) {\bf O}_1(\tau)\;. \nn
\eea
In arriving the first equality we used the fact that $\gamma_1\cO^{\tau}_1=- \cO^{\tau}_1 \gamma_1$ and  $\gamma^2_1=1$. 

   Similarly, we can obtain
\be
U^{\dagger}(\tau)= \mathcal{T}^{\dagger} \;\Pi_{i=1,2} \big(\cosh {\bf O}_i(\tau)+i \gamma_i \sinh {\bf O}_i(\tau) \big)
\ee
where  $ \mathcal{T}^{\dagger}$ denotes the backward time-ordering while $\mathcal{T}$ is the forward one. 

   Using the above and the fact of \eq{topo-c}, we can simplify the reduced density matrix and get
\be \label{reducedSC}
\rho_M(\tau)=C_1C_2\; \rho_{M,0}-S_1C_2\; \sigma_1 \rho_{M,0} \sigma_1-S_2C_1\; \sigma_2 \rho_{M,0} \sigma_2+S_1S_2\; \sigma_3 \rho_{M,0} \sigma_3
\ee
where we have chosen the representation for $\gamma_i$ such that $\gamma_1=\sigma_1$ and $\gamma_2=\sigma_2$ with $\sigma_i$'s the Pauli's matrices, and also defined
\be\label{CiSi}
C_i=\langle \mathcal{T}^{\dagger} \cosh {\bf O}_i (\tau) \mathcal{T} \cosh {\bf O}_i(\tau) \rangle\;, \qquad S_i=\langle \mathcal{T}^{\dagger} \sinh {\bf O}_i (\tau) \mathcal{T} \sinh {\bf O}_i(\tau) \rangle \;.
\ee
Note that, there are no $\langle\mathcal{T}^{\dagger}\sinh\textbf{O}_i(\tau) \mathcal{T} \cosh\textbf{O}_i(\tau)\rangle$ and $\langle\mathcal{T}^{\dagger}\cosh\textbf{O}_i(\tau) \mathcal{T} \sinh\textbf{O}_i(\tau)\rangle$ terms because the expectation values of operators with odd fermion parity are zero.

   We can further simplify $C_i$ and $S_i$ by using the merging formula of OPE, and the details can be found in Appendix A. After doing this, we  arrive the formal result of the reduced dynamics in exactly the same form of \eq{frho-2} and \eq{influence-alpha} as in the static case, except that operators $\cO_i$ are defined by either \eq{OI-1} or \eq{OI-2}.
    
Though we are considering only two Majorana zero modes, it is straightforward to generalize the above formalism to any number of Majorana zero modes. 

\subsection{Environmental spectral densities and ``influence functional"}

    The above formalism is quite generic and can be used for the considerations of topological qubits in the general motions and quadratic couplings. However, for the evaluation of decoherence patterns, in this paper we will the standard bilinear coupling of the Majorana zero mode and the environmental fermi field $\psi$, i.e.,
\be\label{bilinearINT}
\gamma^{\dagger}_i\; \psi_i(t,\vec{x}) + \psi^{\dagger}_i(t,\vec{x})\; \gamma_i
\ee
where $\psi_i(t,\vec{x})$ is a quasi-fermionic field which can be expressed as the following mode expansion form
\be\label{excite-psi}
\psi_i(t,\vec{x})=\lambda_i(\tau_i)\sum_{\vec{k}}\frac{1}{\sqrt{|\vec{k}|}}\big( g_i(\vec{k})\; \hat{a}_{i,\vec{k}}\; e^{-i|\vec{k}| t+i \vec{k}\cdot\vec{x}}+ \tilde{g}_i(\vec{k})\; \hat{b}^\dagger_{i,\vec{k}} \; e^{i|\vec{k}| t-i \vec{k}\cdot \vec{x}}\big)
\ee
where  $\hat{a}_{i,\vec{k}}$ is the annihilation operator for the quasi-particles of spectral weight $g_i(\vec{k})$, and $\hat{b}^\dagger_{i,\vec{k}}$ is the creation operator for the quasi-holes of spectral weight $\tilde{g}_i(\vec{k})$. In the above, $\lambda_i(\tau_i)$'s are the local time-modulation functions depending on local proper time $\tau_i$. 

     By using the facts that the fermionic fields are anti-commuting and the Majorana zero modes are real,  we can rewrite the interaction \eq{bilinearINT} into the form of \eq{intV} with 
\be
\cO_i(t,\vec{x})=\psi_i(t,\vec{x})-\psi^\dagger_i(t,\vec{x})\;.
\ee
One can then follow the formalism presented in section \ref{formalism-time} to obtain the reduced dynamics.

  The interaction \eq{bilinearINT} and \eq{excite-psi} has been adopted in \cite{Zhang} to consider the quantum decoherence of fermionic probe, in which both the probe and environment fermi fields are Dirac type.  In contrast, our probe fermions are Majorana zero modes, which are real and of zero kinetic energy. Besides, our qubit is formed by two spacelike separated Majorana zero modes so that the environmental correlation functions satisfy the locality constraint \eq{topo-c}, but the qubit considered in \cite{Zhang} is made of a single Dirac fermion so that there is no constraint such as \eq{topo-c} imposed. 
  
  Moreover, our interaction \eq{bilinearINT} is in some sense more natural than the coupling to the quadratic environmental observable $\bar{\psi} \psi$ for a usual Unruh-DeWitt monopole detector, e.g. see \cite{Hummer:2015xaa}. If we choose this observable as the environmental operator to which the Majorana zero modes couple, then the interaction Hamiltonian will look like $\gamma_a\gamma_b \bar{\psi}\psi$. This interaction is of higher dimension than our choice \eq{bilinearINT} so that the usual quadratic coupling is more suppressed at low energy and harder to implement in practical. 
\bigskip

  By construction, the spectral density for the excitations produced by $\psi_i$ of \eq{excite-psi} is given by (set the switching function to unity)
\beq
\cA_i(\omega)= \left\{\begin{array}{ll}
\sum_{\vec{k}} \frac{|g_i(\vec{k})|^2}{|\vec{k}|}\delta(\omega-|\vec{k}|), & \mbox{for $\omega>0$} \\ 
\sum_{\vec{k}} \frac{|\tilde{g}_i(\vec{k})|^2}{|\vec{k}|} \delta(\omega+|\vec{k}|), & \mbox{for $\omega<0$}
\end{array} \right.
\eeq     
with the positive-frequency part for the particle excitations and the negative-frequency part from the anti-particle ones.      

    Based on the relation between the spectral density and the Majorana-dressed symmetric Green function as spelled out in Appendix B, we can derive the ``influence functional" for reduced dynamics. Especially, when the Majorana modes are set to motions, the ``influence functional" takes different form for different chosen observers.  This change brings out new features for the reduced dynamics different from the static cases.

    For example, we can choose to observe the reduced dynamics in the comoving frame of  $k$-th Majorana zero mode (we will call this frame ``$M_k$"-frame hereafter, and omit subindex if both Majorana zero modes move in the same way), then the ``influence functional" associated with the $i$-th Majorana zero mode evolves in the proper time $\tau_k$ as follows: 
\bea\label{I-M-0}
\mathcal{I}^{M_k}_i(\tau_k)&=&-2 \int_0^{\tau_k} d\tau_k' \int_0^{\tau_k} d\tau_k'' \;\lambda_i(\tau_i(\tau_k'))\lambda_i(\tau_i(\tau_k''))\int^\infty_{-\infty}d\omega \; |\omega|^{d-1}\cA_i(\omega) e^{-i\omega(t(\tau'_k)-t(\tau_k''))}\nn
\\
&& \quad \times \oint d{\bf \Omega} \exp \Big[ i \omega \hat{n} \cdot (\vec{x}(\tau'_k)-\vec{x}(\tau_k'')) \Big]
\eea
where $\hat{n}:=\vec{k}/|\vec{k}|$ is the unit normal to the unit sphere of measure $\oint d{\bf \Omega}$, and $d$ is the space dimension. Note that the worldline relation \eq{proper-frame} is used implicitly.

   On the other hand, we can also choose to observe the reduced dynamics in the environmental frame (we will call this frame ``E-frame" hereafter) so that the ``influence functional" associated with the $i$-th Majorana zero mode evolves as 
\bea\label{I-E-0}
\mathcal{I}_i^{E}(t)&=&-2 \int_0^t d t' \int_0^t d t''  \;\lambda_i(\tau_i(t'))\lambda_i(\tau_i(t'')){d\tau_i \over dt'}  {d\tau_i \over dt''}  \int^\infty_{-\infty}d\omega \; |\omega|^{d-1} \cA_i(\omega) e^{-i\omega(t'-t'')} \nn
\\
&& \quad \times \oint d{\bf \Omega} \exp \Big[i \omega \hat{n} \cdot [\vec{x}_i(\tau_i(t'))-\vec{x}_i(\tau_i(t''))]\Big]
\eea
where the inverse of the worldline relation \eq{proper-frame} is again used implicitly.

\bigskip

     Later on, we need to input the spectral density when evaluating the reduced dynamics. We will consider four types of (particle-hole symmetric) spectral densities. Below are their explicit forms:
    
\begin{itemize}

\item Uniform spectrum 
\be
\cA(\omega)={q^2 \over |\omega|}\Theta(|\omega|-\Lambda_{IR}):=\cA_{inv}(\omega)
\ee
where $\Theta(x)$ is the Heaviside step function and $\Lambda_{IR}$ is the IR cutoff. Note that $\cA_{inv}(\omega)$ is Lorentz invariant.

\bigskip

The other three types are called the Ohmic type spectrum which takes the form $\mathcal{A}(\omega) \propto |\omega|^Q$. It is called Ohmic, sub-Ohmic and super-Ohmic for $Q=1$, $Q<1$ and $Q>1$, respectively. For each we choose a typical one as follows.

\item Ohmic spectrum 
\be
\cA(\omega)= \frac{q^2}{|\omega|}(\frac{|\omega|}{\Lambda_{UV}})^2 \; e^{-\omega^2/\Lambda_{UV}^2}:=\cA_{ohm}(\omega)
\ee
 
\item Sub-Ohmic spectrum
\be
\cA(\omega)= \frac{q^2}{|\omega|}(\frac{|\omega|}{\Lambda_{UV}})^{\frac{3}{2}}\; e^{-\omega^2/\Lambda_{UV}^2}:=\cA_{sub}(\omega)\;.
\ee

\item Super-Ohmic spectrum
\be
\cA(\omega)= \frac{q^2}{|\omega|}(\frac{|\omega|}{\Lambda_{UV}})^3 \; e^{-\omega^2/\Lambda_{UV}^2}:=\cA_{sup}(\omega)\;.
\ee
\end{itemize}
     In the above $q^2$ and $\Lambda_{UV}$ are some positive constants. Also, for the Ohmic type spectrum we have introduced the factor $e^{-\omega^2/\Lambda_{UV}^2}$ to cutoff the UV divergence. Hereafter, unless specified we will choose $q=1$ for our numerical calcuations. 
          
      Note that in the following numerical calculations,  we will set $\Lambda_{IR}=0.02$ and $\Lambda_{UV}=10$ unless specified.

\subsection{Transition probability from reduced dynamics}\label{section IIIC}
     
        As recently pointed out in \cite{Brenna:2015fga} by studying the transition probability of an Unruh-DeWitt (UDW) detector (an accelerated two-level monopole) \cite{Unruh:1976db,UDW,Crispino:2007eb}, one can find the so-called anti-Unruh phenomena, i.e.,  a UDW detector can cools down as the acceleration increases if it is just accelerated for only a short period. This can be thought as non-equilibrium transient behavior due to the switch-off of the acceleration.  It is interesting to see if the similar phenomenon also occurs for the topological qubit, namely, the Majorana-Unruh-DeWitt (MUDW) detector, and how it will reflect in the decoherence patterns. 
     
   The transition amplitude between the initial state $|{\bf 0}\rangle |i\rangle$ and the final state $|{\bf m}\rangle |f\rangle$ is given by 
\be
A_{i\rightarrow f}^{(\bf m)}:=\lim_{t \rightarrow \infty} \; \langle f| \langle {\bf m}| U(t) |{\bf 0}\rangle |i\rangle
\ee
where $\{ |{\bf m}\rangle \}$ is a complete set of the environmental states, and $|i\rangle$, $|f\rangle$ are the initial and final state of the UDW detector (or the MUDW one), respectively.  Then, the transition probability can be recast into the reduced density matrix as follows:
\be
P_{i\rightarrow f}:=\sum_{\bf m} |A_{i\rightarrow f}^{({\bf m})}|^2=\lim_{\tau \rightarrow \infty} \langle f| \sum_{\bf m} \langle {\bf m}| U(t) |{\bf 0}\rangle |i\rangle \langle i| \langle {\bf 0}| U^{\dagger}(t)  |{\bf m}\rangle |f\rangle=\lim_{\tau \rightarrow \infty}\langle f| \Tr_E \rho^D(t) |f\rangle \;.
\ee
By using \eq{rhoDM}, $\Tr_E \rho^D(t)$ is the reduced density matrix for the (M)UDW detector and note that $\rho^D(t=0)=|{\bf 0}\rangle |i\rangle \langle i| \langle {\bf 0}|$ so that $\rho^D(t)=U(t)\rho^D(t=0) U^{\dagger}(t)$.  

   On the other hand, in the usual treatment for the UDW detector the transition probability is evaluated up to the first order expansion of $U(t)$, see for example \cite{Crispino:2007eb}, i.e.,
\be\label{firstP}
P^{(1)}_{i\rightarrow f}=     \sum_{\bf m} \Big| \int^{\infty} d\tau'   \langle f| \langle {\bf m}| V^{\tau}(\tau')|{\bf 0}\rangle |i\rangle \Big|^2 \;.
\ee  

   Apply the above to the MUDW detector, we have (with $\rho_{00}=1$) the full transition probability 
\be\label{MUDW-P-f}
P_{0\rightarrow 1}=\lim_{t \rightarrow \infty}\; {1\over 2}(1- e^{\mathcal{I}_1(t)+\mathcal{I}_2(t)})
\ee
and for comparison one can evaluate \eq{firstP} explicitly (the details are given in Appendix \ref{app C}) and get 
\be\label{MUDW-P-1}
P^{(1)}_{0\rightarrow 1}=\lim_{t \rightarrow \infty} -{1\over 2} \Big( \mathcal{I}_1(t)+\mathcal{I}_2(t)\Big)
\ee   
which agrees with the first order expansion of \eq{MUDW-P-f}.  

 Note that the above results apply for different reference frames by just changing the ``influence functionals" $\mathcal{I}_i(t)$ accordingly.  Moreover, one can also relax the condition $t\rightarrow \infty$ to obtain the instantaneous transition probability at any time.
     
      In most of the cases, the MUDW will decohere so that the final state is ``classical" and the full transition probability is zero.  In such cases, the first order result \eq{MUDW-P-1} is ill-defined as $\mathcal{I}_i(t\rightarrow\infty) \rightarrow -\infty$. From our formulation, it is clear that this divergence is due to the truncation of the higher order terms as the full transition probability is well-defined. There are two ways of remedy. One way is instead to consider the transition rate defined by
\be
\mathcal{R}_{0\rightarrow 1}:= \lim_{T \rightarrow \infty}  { \sum_m  \Big| \int^T d\tau' \ \langle f| {}_e\langle m| V^{\tau}(\tau')|0\rangle_e |i\rangle \Big|^2  \over T} \;.     
\ee
This will yield a finite result, and is usually adopted to derive the Unruh effect. The other way is to introduce time-modulation function $\lambda_i(\tau)$ of finite duration. In this way, the decoherence will stop after the interaction is turned off so that the transition probability can be nonzero and well-defined. The short time modulation causes some nontrivial transient effect for the accelerating UDW detector, dubbed as ``anti-Unruh" phenomenon  \cite{Brenna:2015fga}.  In the next section, we will see this ``anti-Uuruh" phenomenon indeed yield nontrivial decoherence patterns. 

    In the next section, we will see that the constantly-accelerating topological qubit will always decohere completely into Gibbs state, i.e., $\mathcal{I}_i(t\rightarrow\infty) \rightarrow -\infty$ so that  
\be  \label{finalGibbs}
\rho_M(t\rightarrow \infty)=\left(\begin{array}{cc}{1\over 2} \;
& \; 0 \\ 0 \; &  {1\over 2} \end{array}\right)\;,
\ee
even for very small acceleration.  Recall that the topological qubit has degenerate zero energy spectrum as the Majorana modes are zero modes. Therefore, if we think the topological qubit is thermalized due to the Unruh effect, then the final state of topological qubit will turn into Gibbs state \eq{finalGibbs} which is independent of acceleration due to the degenerate zero energy spectrum.

\section{Reduced dynamics of topological qubit in linear motions} \label{sec IV}
  
   In this section we will consider the decoherence patterns of topological qubit in linear motions, see the left panel of Fig. \ref{cartoon}, where two Majorana zero modes $\gamma_{1,2}$ form a topological qubit. As shown in \eq{frho-2}, the reduced dynamics is controlled by the ``influence functionals" associated with each Majorana modes, we will then evaluate them in both the $M_k$-frame and $E$-frame for each specific linear motion trajectory.

\subsection{``Overtaking" phenomenon: Constant acceleration} \label{overtaking}

   Before considering the generic motion of the topological qubit, we start with the case of constant acceleration. Then, it is natural to ask if the reduced dynamics will be the finite-temperature version of the static topological qubit as expected by the Unruh effect.

   The worldline of a particle with constant acceleration $a$ is given by 
\be\label{constant-a-wl}
t(\tau)=\frac{\sinh a\tau}{a},\qquad x(\tau)=\frac{\cosh a\tau-1}{a}\;.
\ee
We assume this trajectory holds for both Majorana zero modes of the topological qubit.  Then, by evaluating the ``influence functional" \eq{I-M-0} in the $M$-frame, we obtain
\be
\mathcal{I}^{M}(\tau)=-\frac{1}{a^2}\int_{-\infty}^\infty  d\omega \; \mathcal{A}(\omega) \big(|\textrm{E}_1(i\frac{\omega}{a})-\textrm{E}_1(i\frac{\omega}{a}e^{-a\tau})|^2+|\textrm{E}_1(-i\frac{\omega}{a})-\textrm{E}_1(-i\frac{\omega}{a}e^{a\tau})|^2 \big)   
\ee
where  
\be
\textrm{E}_1(z):=\int_1^\infty \frac{e^{-z u}}{u}du\; ;
\ee
or by evaluating \eq{I-E-0} in the $E$-frame we get
\be
\mathcal{I}^E(t)=-\frac{1}{a^2}\int_{-\infty}^\infty  d\omega \; \mathcal{A}(\omega) \big(|\textrm{E}_1(i\frac{\omega}{a})-\textrm{E}_1(i\frac{\omega}{a}e^{-a\tau(t)})|^2+|\textrm{E}_1(-i\frac{\omega}{a})-\textrm{E}_1(-i\frac{\omega}{a}e^{a\tau(t)})|^2 \big)   
\ee
where
\be
\tau(t)=\frac{\sinh^{-1}at}{a}\;.
\ee
In the above we have assumed the time-modulation functions to be unity (time-independent) and also omitted the Majorana index as both zero modes move in the same way.
         
    On the other hand, applying Wick's theorem as shown in \cite{Evans} we can obtain the reduced dynamics in the thermal environment by just replacing the Green functions with their thermal versions. In our case, the resultant ``influence functional" should look like
\be
\mathcal{I}^{Thermal}(t)=-8\int_{-\infty}^\infty d\omega \; [1+2n(\omega)]\textrm{sgn}(\omega) \mathcal{A}(\omega) \;\frac{1-\cos \omega t}{\omega^2} 
\ee
where $n(\omega)$ is the Bose-Einstein distribution with temperature ${a\over 2\pi}$ by recalling that our Majorana-dressed Green functions are bosonic defined by the time-ordering when applying Wick's theorem.

\begin{figure}
\includegraphics[width=.5\columnwidth]{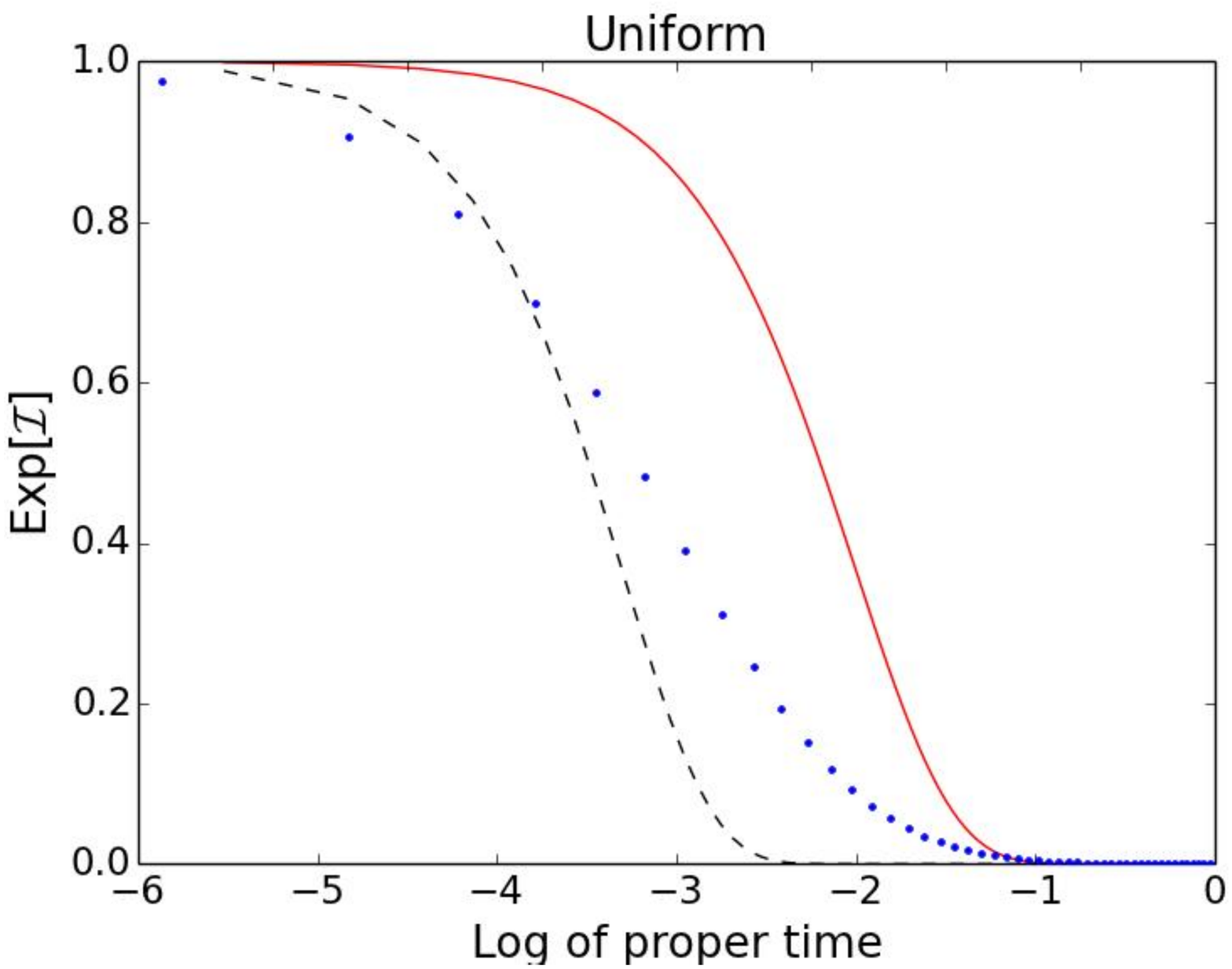}
\includegraphics[width=.5\columnwidth]{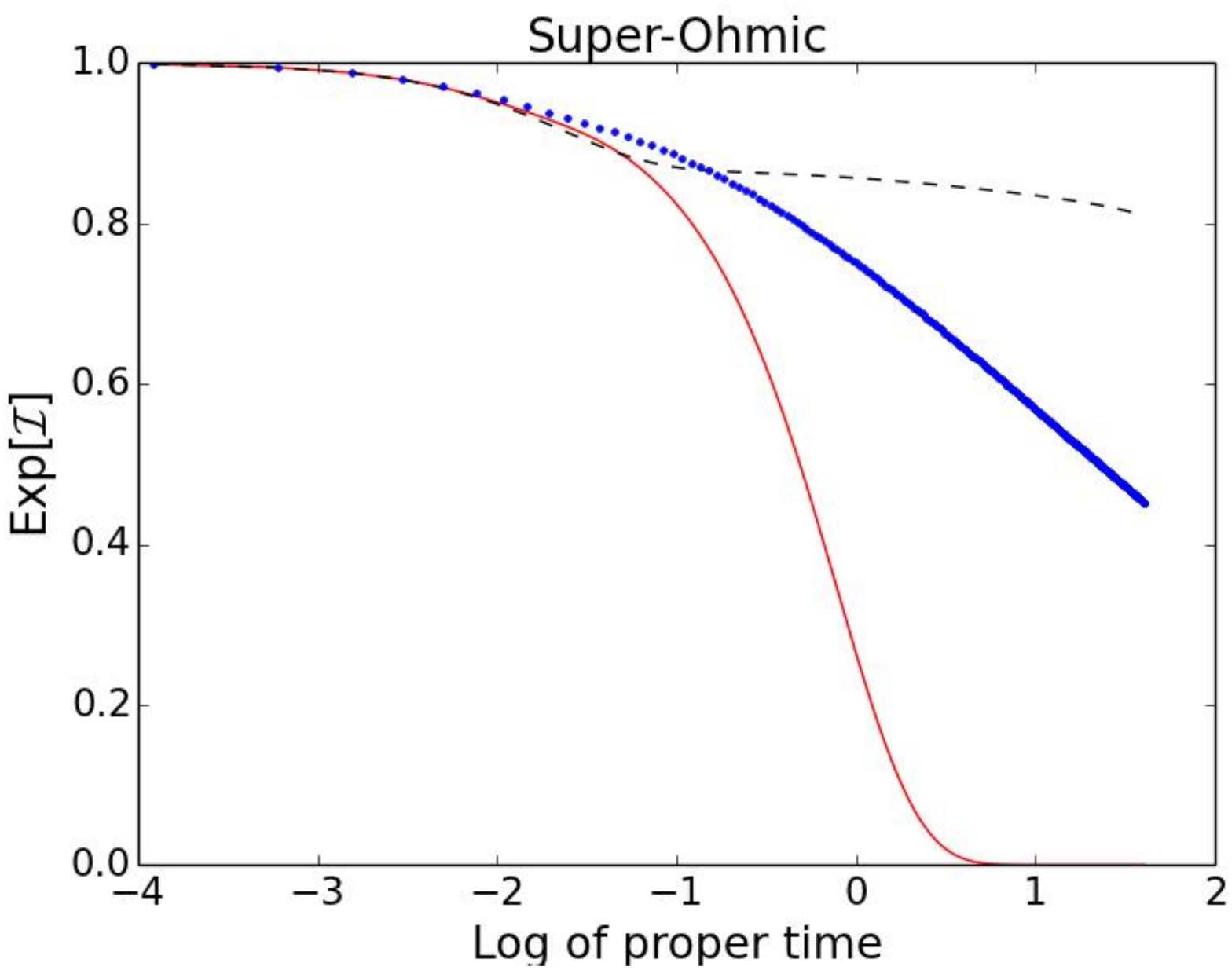}
\caption{ Decoherence patterns (exponential of ``influence functionals" vs log of proper time) of a constantly-accelerating topological qubit in various frames in the environments of uniform spectrum (Left) and of super-Ohmic spectrum (Right):  in the $M$-frame (solid red), in the $E$-frame (dotted blue) and of the thermal one (dashed black) with the acceleration $a=5$. } 
\label{const-thermal}
\end{figure}

     By comparing the above ``influence functionals" (e.g., in the environments of uniform and super-Ohmic spectrum, respectively), we can see that the reduced dynamics of a constantly-accelerating topological qubit is different from the one in the thermal environment.  As the decay rates of the off-diagonal elements of the reduced density matrix, see \eq{frho-2}, are dictated by the ``influence functional", this comparison will demonstrate the difference of the decoherence patterns. The results are explicitly shown in Fig. \ref{const-thermal} where we plot the exponential of the ``influence functional" versus the log of chosen proper time in $E$-frame, $M$-frame and thermal environment. 
     
     It is clear that the decoherence patterns of the accelerating topological qubit in either $E$- or $M$-frames are quite different from the ones of the static qubit in the corresponding thermal environments.  The difference could be attributed to either non-Markovinity or the non-local nature of the topological qubit. To pin down the exact reason one should study the decoherence patterns of the usual local qubit. This should be an interesting problem for the future work.  Moreover, due to the time-dilation effect, the decoherence rate in the $E$-frame is smaller than the one in the $M$-frame as shown in Fig. \ref{const-thermal}.

\begin{figure} 
\includegraphics[width=.5\columnwidth]{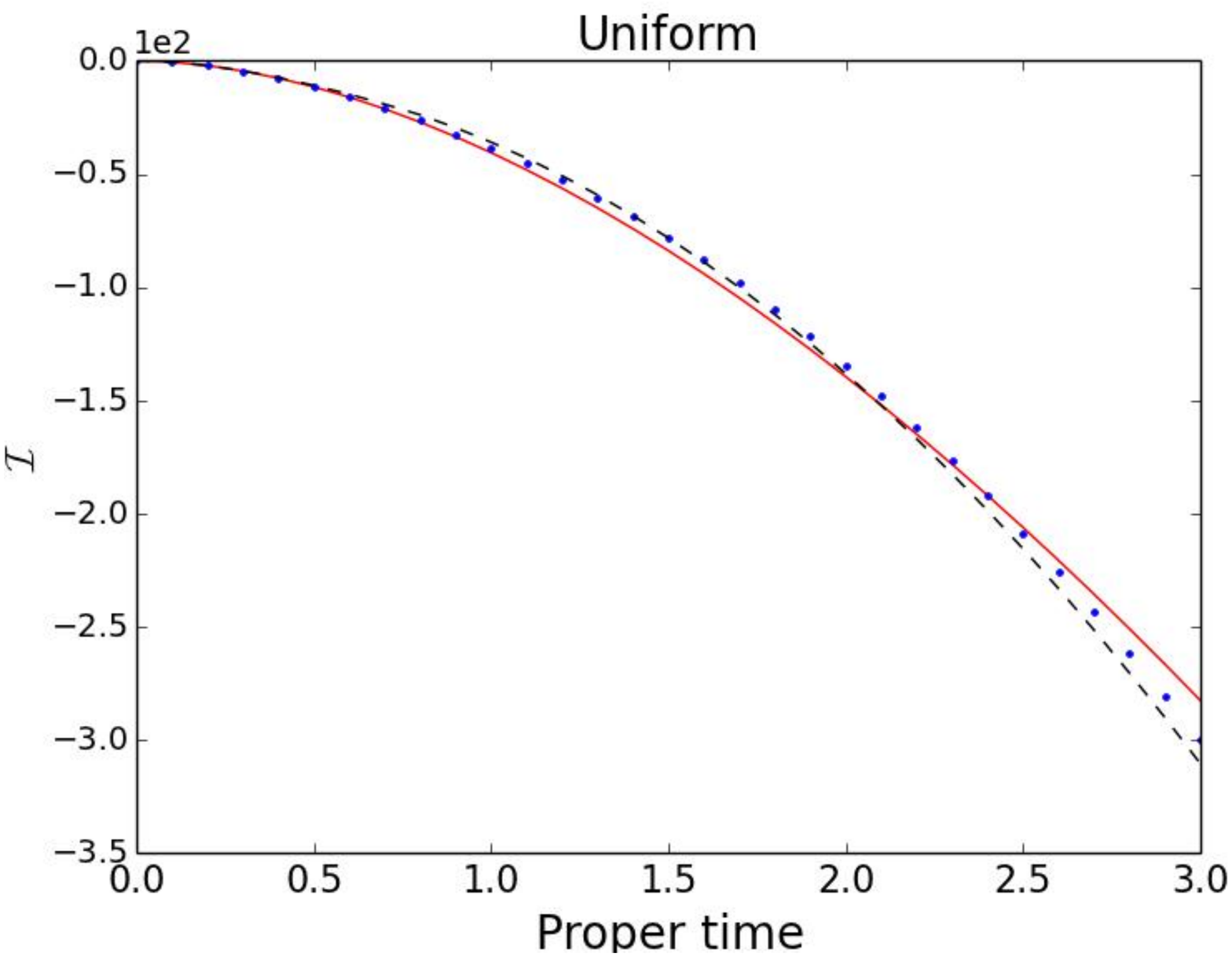}
\includegraphics[width=.5\columnwidth]{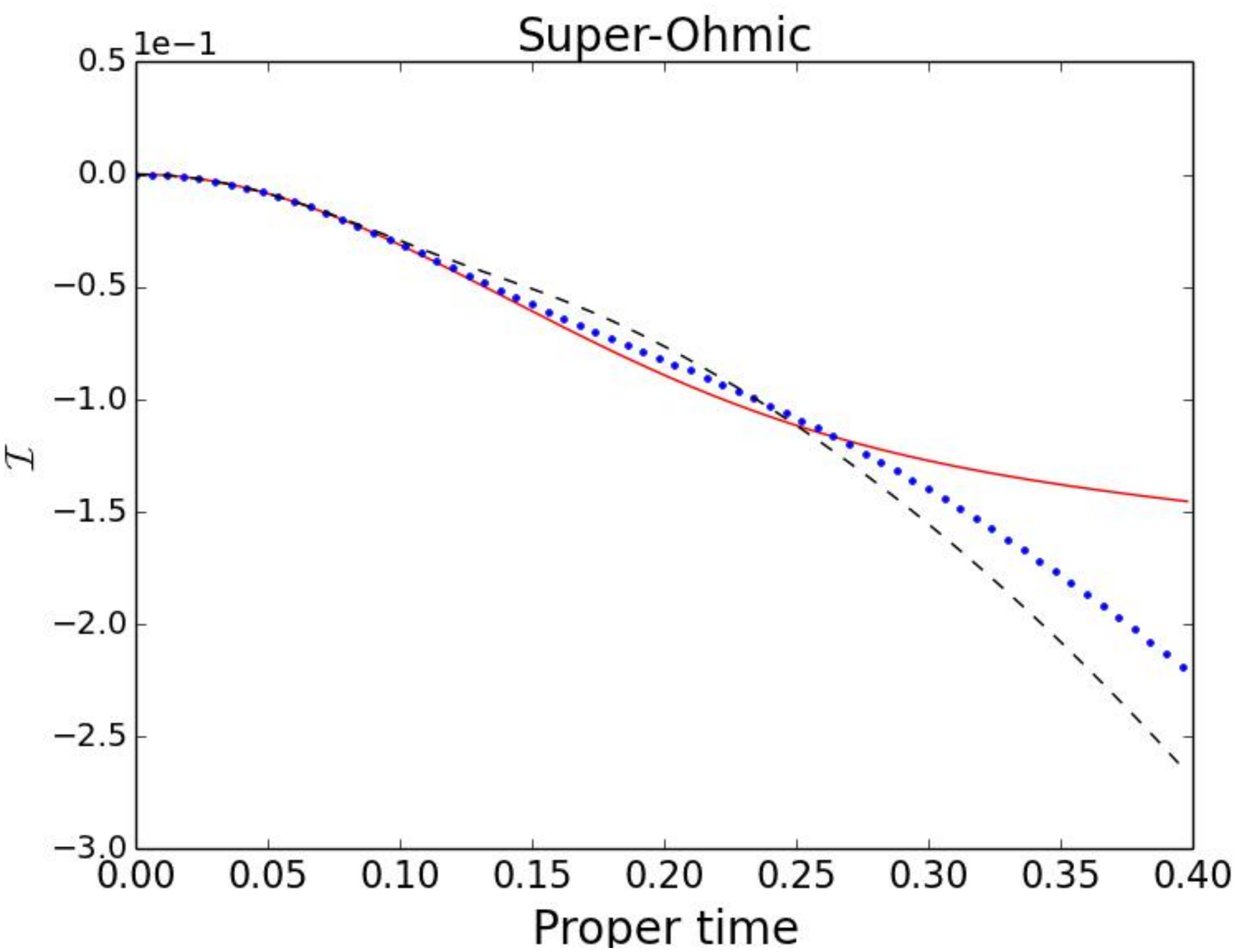}
\caption{``Influence functionals" of a constantly-accelerating topological qubit in the $M$-frame in the environments of uniform spectrum (Left) and of super-Ohmic spectrum (Right) for different accelerations: $a=1$ (solid red), $a=5$ (dotted blue) and $a=10$ (dashed black). Note that the decoherence patterns show the ``overtaking" phenomenon. } 
\label{const-a}
\end{figure}

     Next, we focus on the decoherence in the $M$-frame by tuning the magnitude of the acceleration. We show the results for the environments of uniform and super-Ohmic spectrum in Fig. \ref{const-a}.  Note that here we plot  the ``influence functionals" instead of their exponentials to magnify the details of the so-called ``overtaking" phenomenon (especially in the environment of uniform spectrum) discussed below.

     Firstly, we find that in all cases the topological qubit will completely decohere into Gibbs state \eq{finalGibbs} in the long run. Recall that the static topological qubit is robust against decoherence in the super-Ohmic environment \cite{MajDeco}. Our results imply that the accelerating topological qubit is completely thermalized due to Unruh effect even though the ``influence functionals" are different from the thermal ones. However, the final Gibbs state has no acceleration dependence due to the degenerate zero energy spectrum of topological qubit as commented around \eq{finalGibbs}.  That is, the transition probability $P_{0\rightarrow 1}$ \eq{MUDW-P-f} will be equal to $1/2$.

     Besides, from Fig. \ref{const-a}  we observe an interesting phenomenon which we call ``overtaking": the topological qubit with smaller acceleration decoheres/thermalizes faster at beginning  but then slower than the one with larger acceleration in the late time. Note that the ``overtaking" with different accelerations occurs almost around the same time.  
     
     Moreover, as seen from Fig. \ref{const-a},  the magnitudes of the ``influence functionals" $\mathcal{I}$ in the environments of uniform spectrum are far larger than  in the super-Ohmic environments, so that the corresponding $e^{\mathcal{I}}$'s are far smaller as $\mathcal{I}$'s are negative.  This implies that the ``overtaking" is far less obvious in the environments of uniform spectrum than in the Ohmic type  environments, e.g. the super-Ohmic one as shown in Fig. \ref{const-a}. This is also the reason why we plot $\mathcal{I}$ instead of $e^{\mathcal I}$ to magnify the details of the ``overtaking" in the cases of uniform spectrum. \footnote{We have checked that the ``overtaking" is numerically stable against the tuning of time-step size (from 0.002 to 0.01) when numerically evaluating the $\mathcal{I}$. }  The almost-``overtaking"-free feature in the environment of the uniform spectrum could be due to its invariance under Lorentz transformations.

\subsection{``Overtaking" phenomenon: Constant velocity}

   Next, we turn into an even simpler setting but interesting enough to compare with the constant acceleration cases: the boost topological qubit with constant velocity. In this case, the worldline of the Majorana zero modes with constant velocity $v$ (in (1+1)D spacetime) is given by 
\be \label{constv-wl}
t(\tau)=\frac{\tau}{\sqrt{1-v^2}},\qquad x(\tau)=\frac{v\tau}{\sqrt{1-v^2}}\;.
\ee

\begin{figure}
\includegraphics[width=.5\columnwidth]{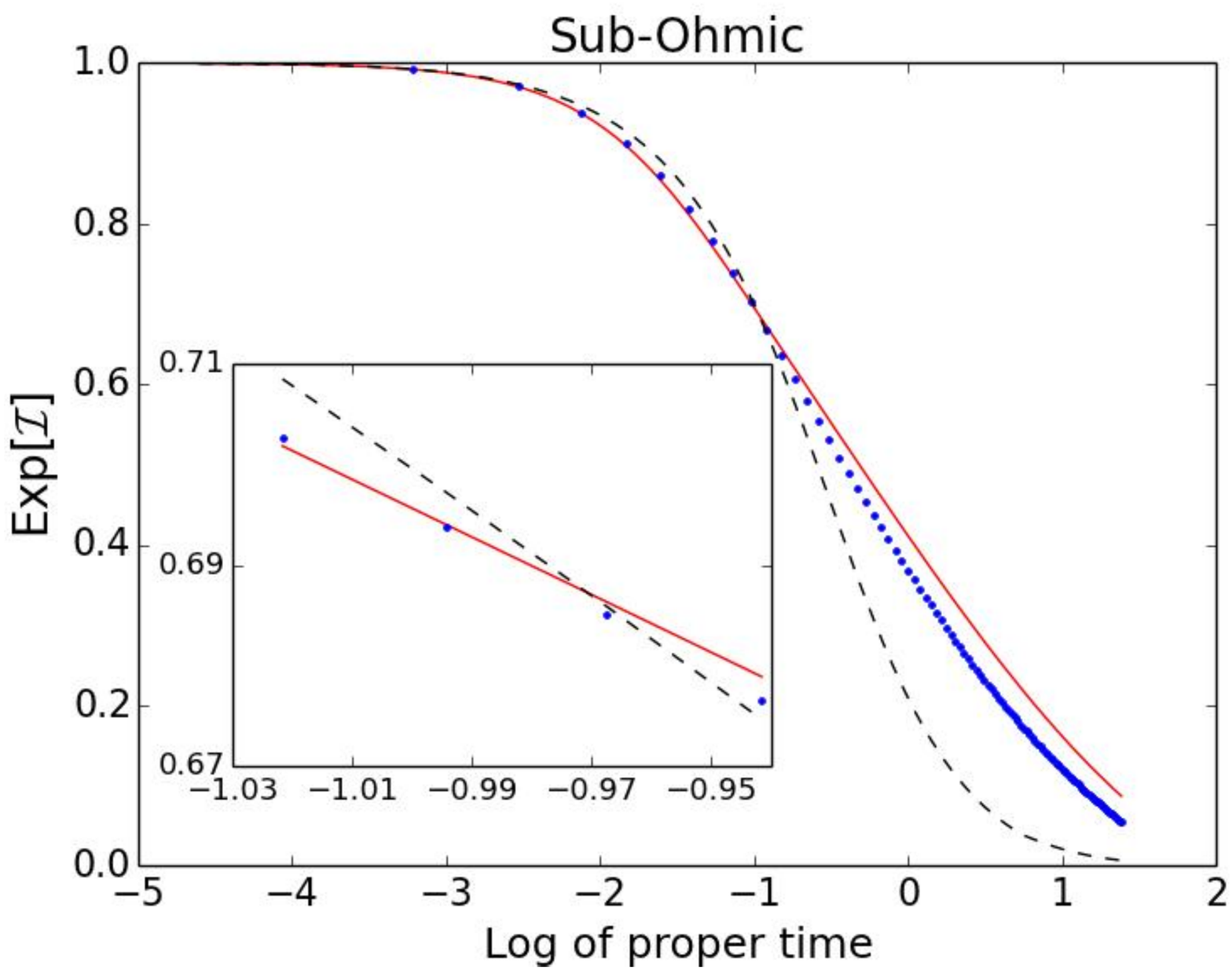}
\includegraphics[width=.5\columnwidth]{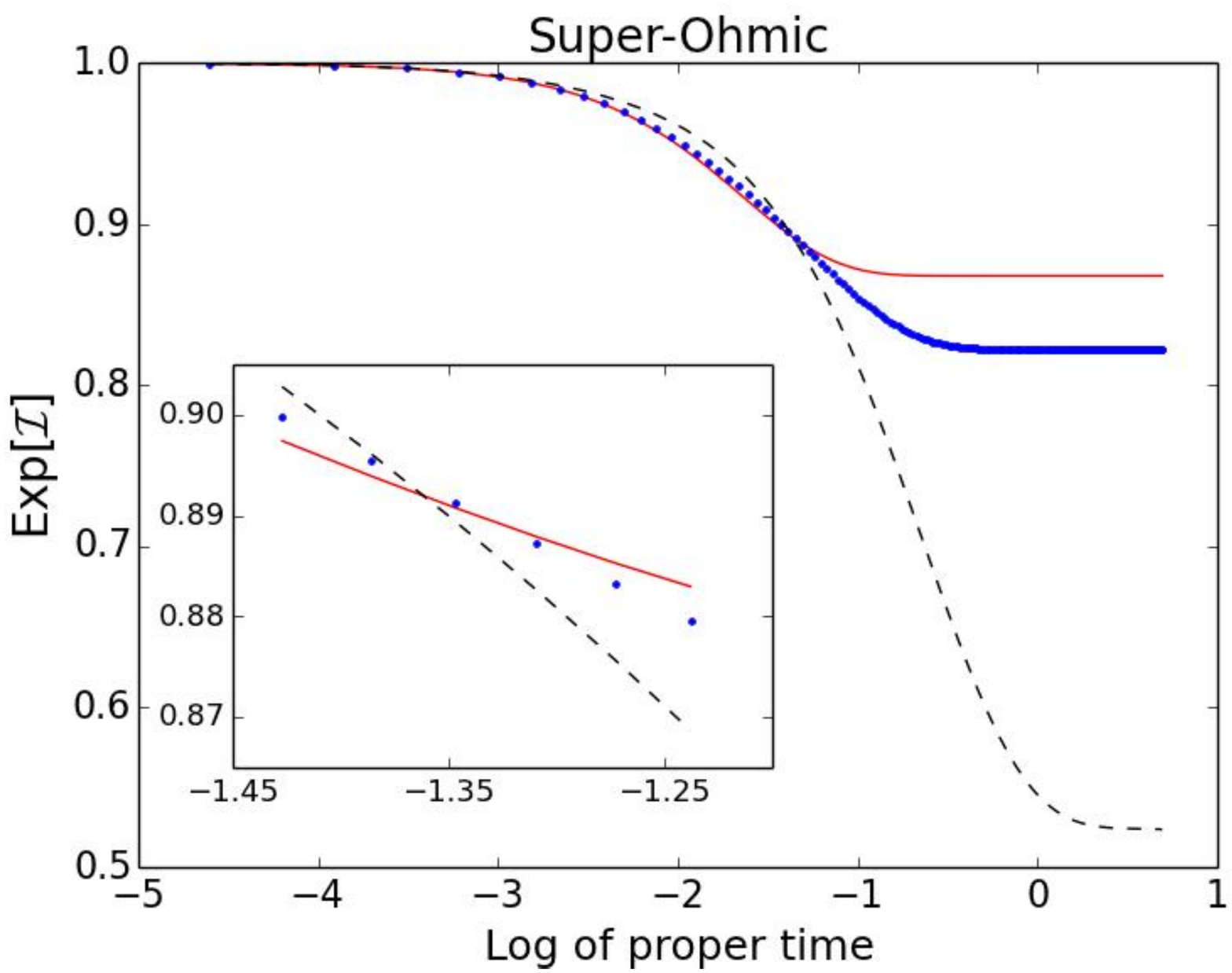}
\caption{Decoherence patterns of a boost topological qubit in the $M$-frame in the environments of sub-Ohmic spectrum (Left) and of super-Ohmic spectrum (Right) for different velocities: $v=0$ (solid red), $v=0.4$ (dotted blue) and $v=0.8$ (dashed black). The insets show the details of ``overtaking" phenomena during the decohrence.} 
\label{const-v}
\end{figure}   

     As the uniform spectrum is Lorentz invariant, so is the corresponding ``influence functional". Therefore, the decoherence pattern in the environment of uniform spectrum will be the same for all the Lorentz boosts of the topological qubit. However, this is not the case for the Ohmic type environments and the corresponding decoherence patterns with both Majorana zero modes moving along the worldline \eq{constv-wl} are shown in Fig.\ref{const-v}.   For the sub-Ohmic environment, the boost only change the rates of the decoherence and will not change the final state, i.e., it completely thermalize into Gibbs state. On the other hand, for the super-Ohmic environment, the final state does change due to the boost but it will not thermalize completely. This implies that there is no Unruh effect as expected. 
     
     Moreover, ``overtaking" occurs again. Similarly, in this case, the slower topological qubit decoheres/thermalizes faster at the beginning but then slower in the late time than the fast-moving one.

      Combined with the results for the cases with constant acceleration, it implies that the ``overtaking" phenomenon could mainly be due to the breaking of Lorentz invariance as the ``overtaking" phenomenon in the uniform-spectrum environment is (null or) far less obvious than the ones in the Ohmic type environments.

\subsection{Decoherence Impedance and ``Anti-Unruh"}

       The ``overtaking" phenomenon suggests that the quantum system has some kind of ``impedance" against decoherence when subjecting to the sudden change of external conditions such as boost or acceleration. That is, the more violent change will instead cause slower decoherence at the very beginning. However, finally the thermalization takes over and the ``impedance" effect will be washed out. This is why the ``overtaking" occurs as the thermalization finally overcomes the {\bf decoherence impedance}. 
            
        This kind of decoherence impedance reminds us the ``anti-Unruh" phenomenon recently studied in \cite{Brenna:2015fga}. In this subsection we will explore the relation between them in the setup of our MUDW detector. 
 
     In the usual UDW setup for the Unruh effect, the higher acceleration will result in higher temperature so that the transition probability of the UDW detector goes up. However, if one decouple the UDW detector from the environment after a finite duration, an interesting phenomenon coined as ``anti-Unruh" can be observed from the transition probability of the UDW detector: the transition probability decreases as the acceleration increases if the coupling of the detector with the environmental field lasts only for a short enough finite duration. \cite{Brenna:2015fga}.  Namely, the coupling constant is a function of time, e.g. 
\be\label{sigma-duration}
\lambda(\tau)= e^{-\frac{\tau^2}{\sigma^2}}
\ee
where $\sigma$ is the duration scale.  For small enough $\sigma$, the ``anti-Unruh" will occur.  
     
   As in our MUDW setup we have the exact expression for the first-order and full transition probabilities, i.e., \eq{MUDW-P-1} and \eq{MUDW-P-f}, respectively, we can check if we can observe the similar ``anti-Unruh" phenomenon  by tuning $\sigma$. Note that the transition probability of MUDW detector is dictated by the ``influence functionals". 
   
   On the other hand, the ``overtaking" or the decoherence impedance shown in the decoherence patterns are also dictated by the ``influence functionals". Thus, we shall expect that the decoherence impedance and the ``anti-Unruh" shall imply each other.   The decoherence impedance appears in the early stage of the decoherence patterns, i.e., the larger acceleration causes slower decoherence. This must correspond to the ``anti-Unruh" for the small duration coupling.

\begin{figure}
\includegraphics[width=.5\columnwidth]{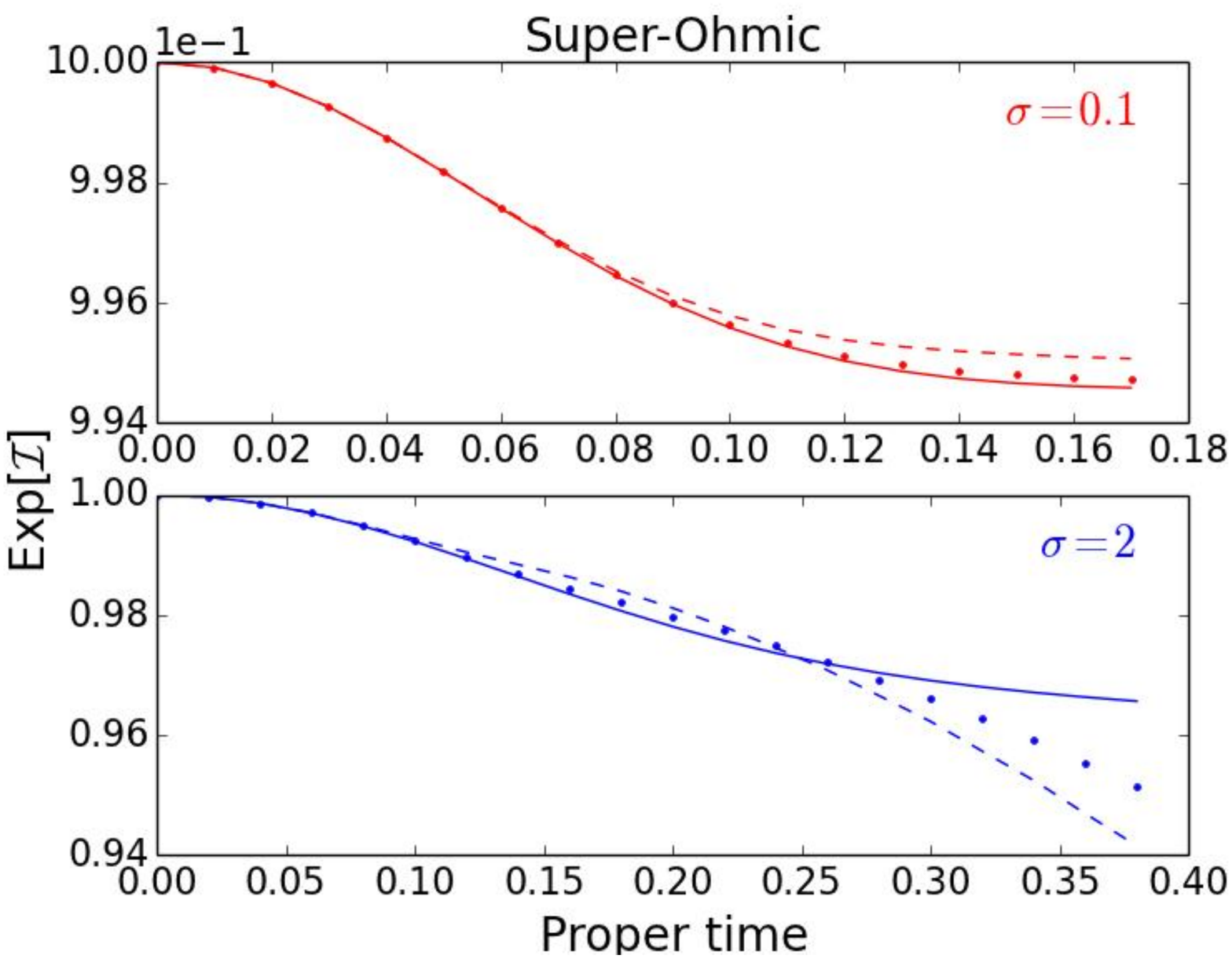}
\includegraphics[width=.5\columnwidth]{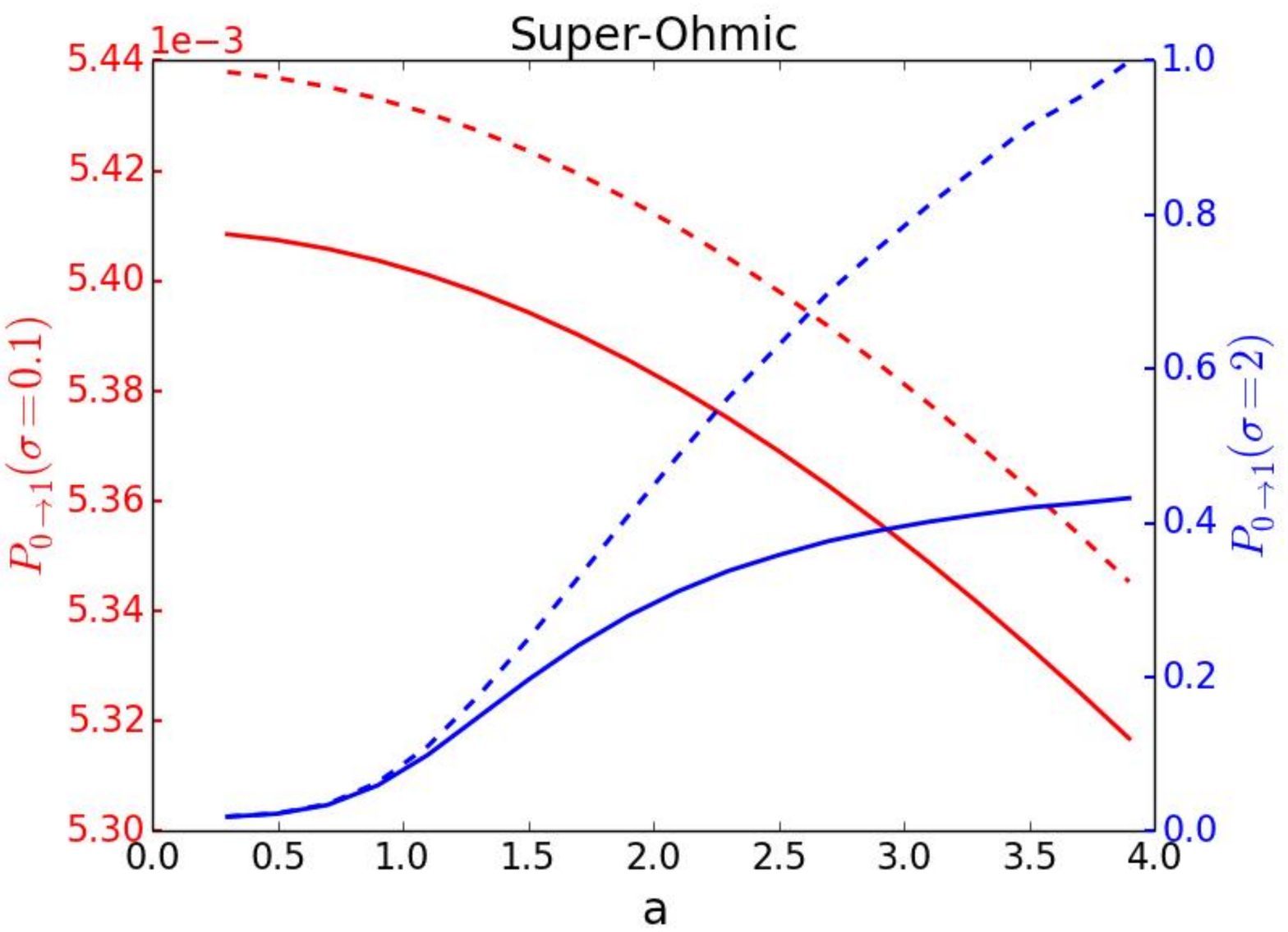}
\caption{Decoherence patterns and transition probabilities in the $M$-frame of a constantly accelerating MUDW detector in the super-Ohmic environment (with $q=0.5$) with the switching function of the time duration scales: $\sigma=0.1$ (red) and $\sigma=2$ (blue). Left : Decoherence patterns with $a=1$ (solid), $a=5$ (dashed) and $a=10$ (dotted). Right: Transition probabilities $P_{0\rightarrow 1}$ (solid) and $P^{(1)}_{0\rightarrow 1}$ (dashed) versus acceleration $a$. The left y-axis is for $P_{0\rightarrow 1}$ and  $P^{(1)}_{0\rightarrow 1}$ with $\sigma=0.1$, and the right y-axis with $\sigma=2$. This figure shows that the ``overtaking" and ``anti-Unruh" imply each other.} 
\label{sigma-a}
\end{figure}

   To demonstrate this possible connection, in Fig. \ref{sigma-a} we juxtapose the decoherence pattern and the transition probability of MUDW detector in the super-Ohmic environment. We choose the Ohmic type environment here as it has more obvious ``overtaking" than the one of uniform spectrum.   In the left panel we plot the decoherence patterns for small time duration ($\sigma=0.1$) and the large one ($\sigma=2$). One can see that only the latter shows the ``overtaking" so that the initial effect of decoherence impedance is washed out by the thermalization. We shall then expect the ``anti-Unruh" shall only occur for the $\sigma=0.1$ one as the decoherence impedance is still in charge. This is indeed the case as shown in the right panel in which we plot the corresponding transition probabilities which show the ``anti-Unruh" only for the $\sigma=0.1$ case. Note that we have plotted both the first-order and the full transition probabilities, and they share the same qualitative features.  The results show that the ``overtaking" and ``anti-Unruh" imply each other, as speculated above.

\begin{figure}
\includegraphics[width=.5\columnwidth]{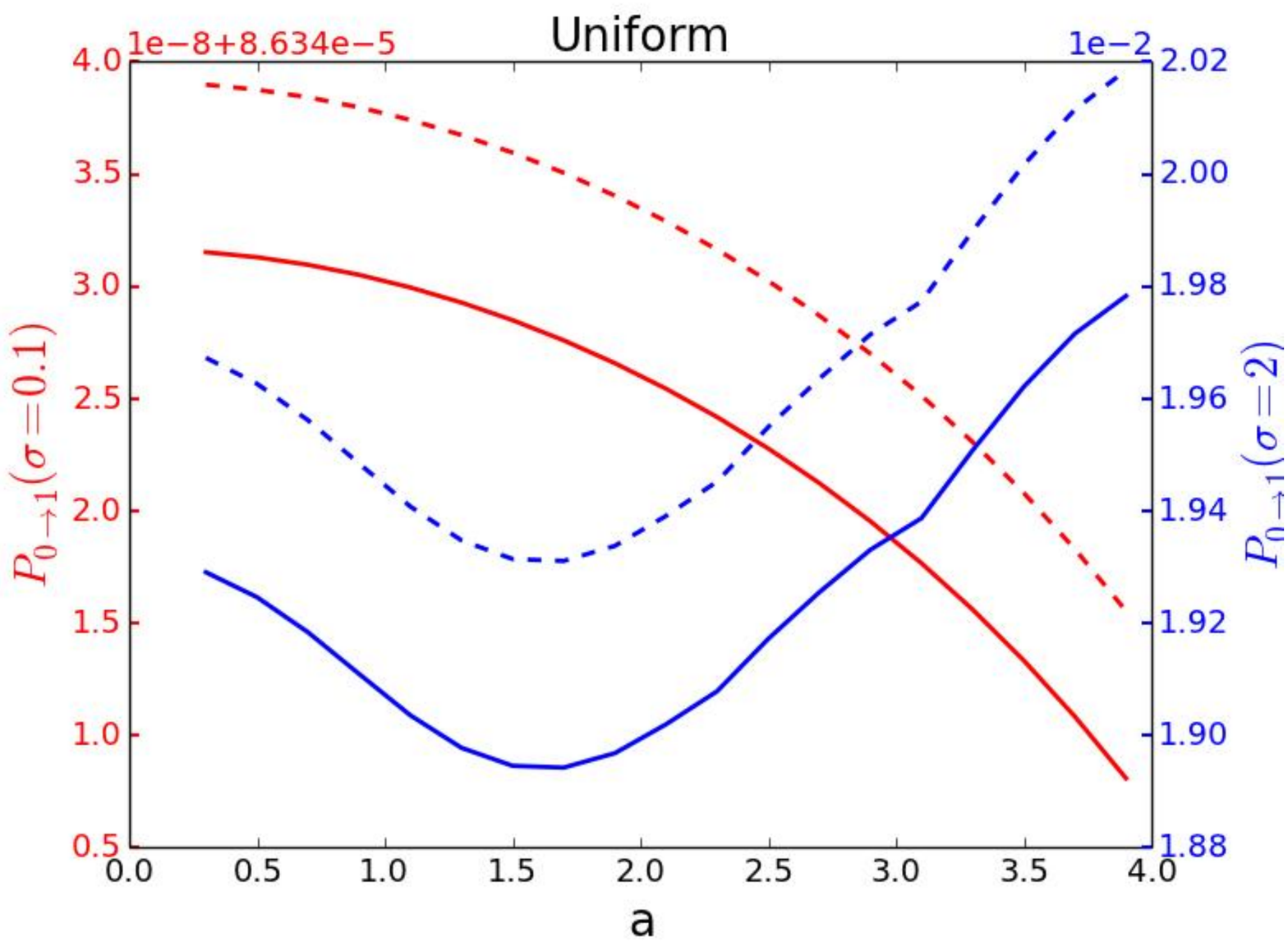}
\includegraphics[width=.5\columnwidth]{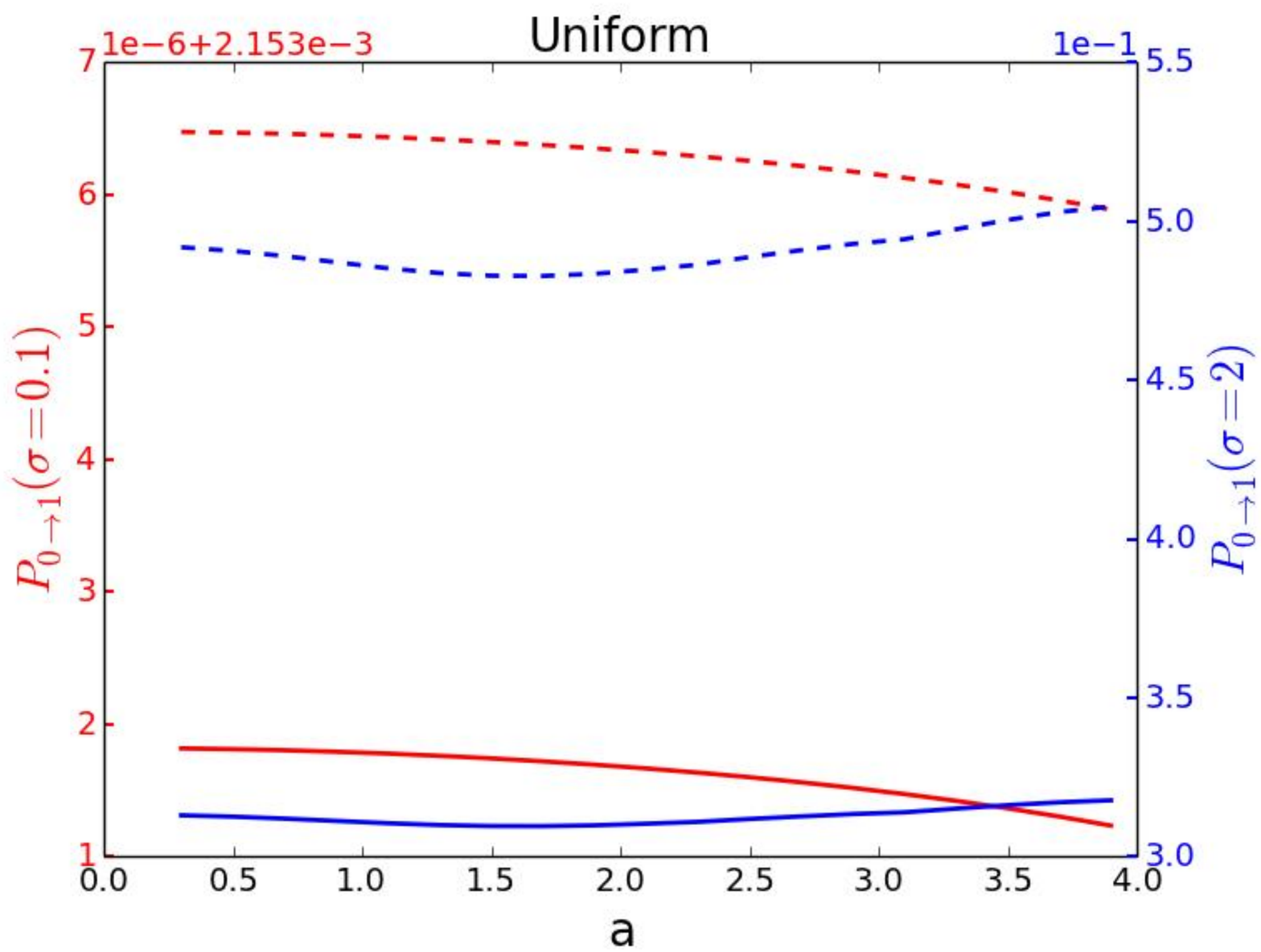}
\caption{Transition probabilities $P_{0\rightarrow 1}$ (solid) and $P^{(1)}_{0\rightarrow 1}$ (dashed)  in the $M$-frame of a constantly accelerating MUDW detector with the switching function of the time duration scales:  $\sigma=0.1$ (red) and $\sigma=2$ (blue)  in the environments of uniform spectrum with $q=0.01$ (Left) and $q=0.05$ (Right). The IR cutoff $\Lambda_{IR}=0.02$.} 
\label{trans-uniform-aU}
\end{figure}

\begin{figure}
\includegraphics[width=1. \columnwidth, height=.5 \columnwidth]{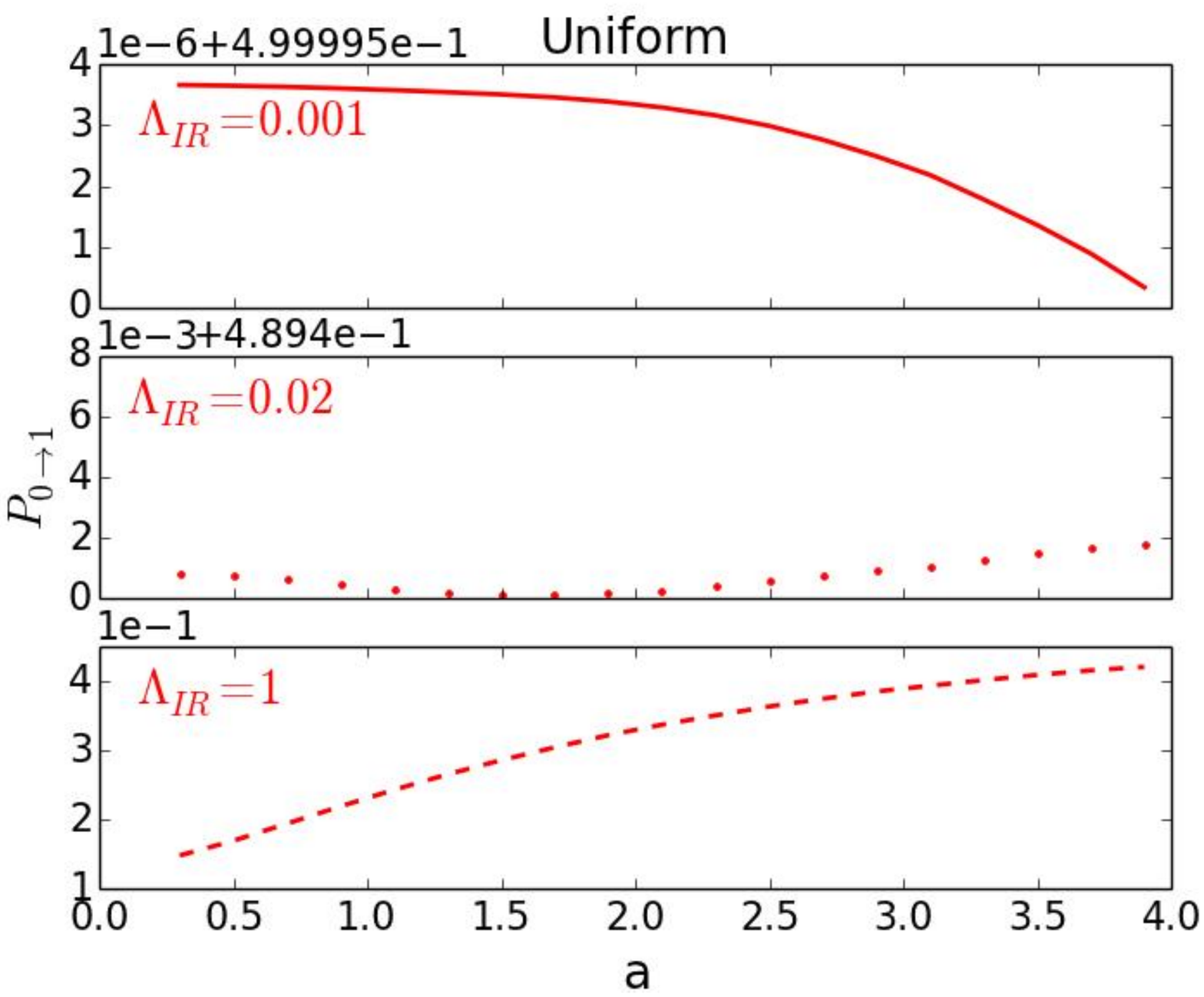}
\caption{Transition probabilities $P_{0\rightarrow 1}$ in the $M$-frame of a constantly accelerating MUDW detector with a switching function of time duration scale $\sigma=2$  in the environments of uniform spectrum with $q=0.01$ and the IR cutoffs: $\Lambda_{IR}=0.001$ (solid),  $\Lambda_{IR}=0.02$ (dotted) and  $\Lambda_{IR}=1$ (dashed).} 
\label{IR-aU}
\end{figure}

    Note that the thermalization goes faster if the strength $q$ of the environmental spectrum is larger. Thus, even with fixed time duration scale, if $q$ is large enough the thermalization shall always dominate the decoherence impedance. Our numerical check confirms this, i.e., the full transition probability $P_{0\rightarrow 1}$  equal to $1/2$ almost for all accelerations.  The same happens if we tune $\sigma$ to be large enough with $q$ fixed.

   Now we turn to the more peculiar ``anti-Unruh" patterns for the environments of uniform spectrum by tuning $\sigma$, $q$ as shown in Fig. \ref{trans-uniform-aU} and by tuning $\Lambda_{IR}$ as shown in Fig. \ref{IR-aU}.  In Fig. \ref{trans-uniform-aU} both $P_{0\rightarrow 1}$ and $P^{(1)}_{0\rightarrow 1}$ show a crossover from the ``anti-Unruh" to the usual Unruh pattern as the acceleration increases for the $\sigma=2$ cases no matter of $q=0.01$ or $q=0.05$.   On the other hand, this peculiar crossover disappears for the  case with $\sigma=0.1$ but just remains the anti-Unruh pattern.  Similar peculiar crossover appears when we tune the IR cutoff $\Lambda_{IR}$ as shown in Fig. \ref{IR-aU}.  Unlike the Ohmic type environment, the uniform spectrum has an IR cutoff dependence. From the results in Fig. \ref{trans-uniform-aU} and Fig. \ref{IR-aU}, this peculiar crossover could be due to interplay of tuning $\Lambda_{IR}$, $\sigma$ and $q$  so that it may not be generic.

\subsection{Information backflow and time modulation}

   The important feature of our reduced dynamics is that it was solved exactly without making Markov approximation. Thus, we shall expect the non-Markovian decoherence patterns of the topological qubit, which will be characterized mainly by the information backflow. There have been issues of quantifying the non-Marokovinity in the past few years, see \cite {ZhangPRL,BreuerPRL,PlenioPRL} for examples. However, here we will concern only the qualitative but not the quantitative feature of non-Markovinity/information backflow, which is characterized by the non-monotonic decoherence patterns, i.e., the off-diagonal matrix elements of the reduced density matrix do not decay monotonically in time.

   Moreover, as the Majorana zero modes are degenerate with zero energy, there is no energy flow but only information flow between the topological qubit and the environment during the decoherence. This may in contrast to the case of usual qubit/probe with non-zero kinetic energy for which the resultant non-monotonic decoherence patterns may also be assisted by the energy backflow.

   In the following we will tune the time dependence of either the switching function or the acceleration to see when such behaviors of information backflow will be still preserved.

\subsubsection{Parameter space for information backflow}

  In \cite{MajDeco}, it showed that the static topological qubit in the super-Ohmic environment does show the non-Markovian behavior characterized in the way just described. As we expect the thermalization will damp the information backflow, it is then interesting to explore this possible connection from the decoherence patterns of the topological qubit in the super-Ohmic environment by tuning the parameters controlling the degrees of the thermalization. 
  
\begin{figure}
\includegraphics[width=1.\columnwidth, height=.5 \columnwidth]{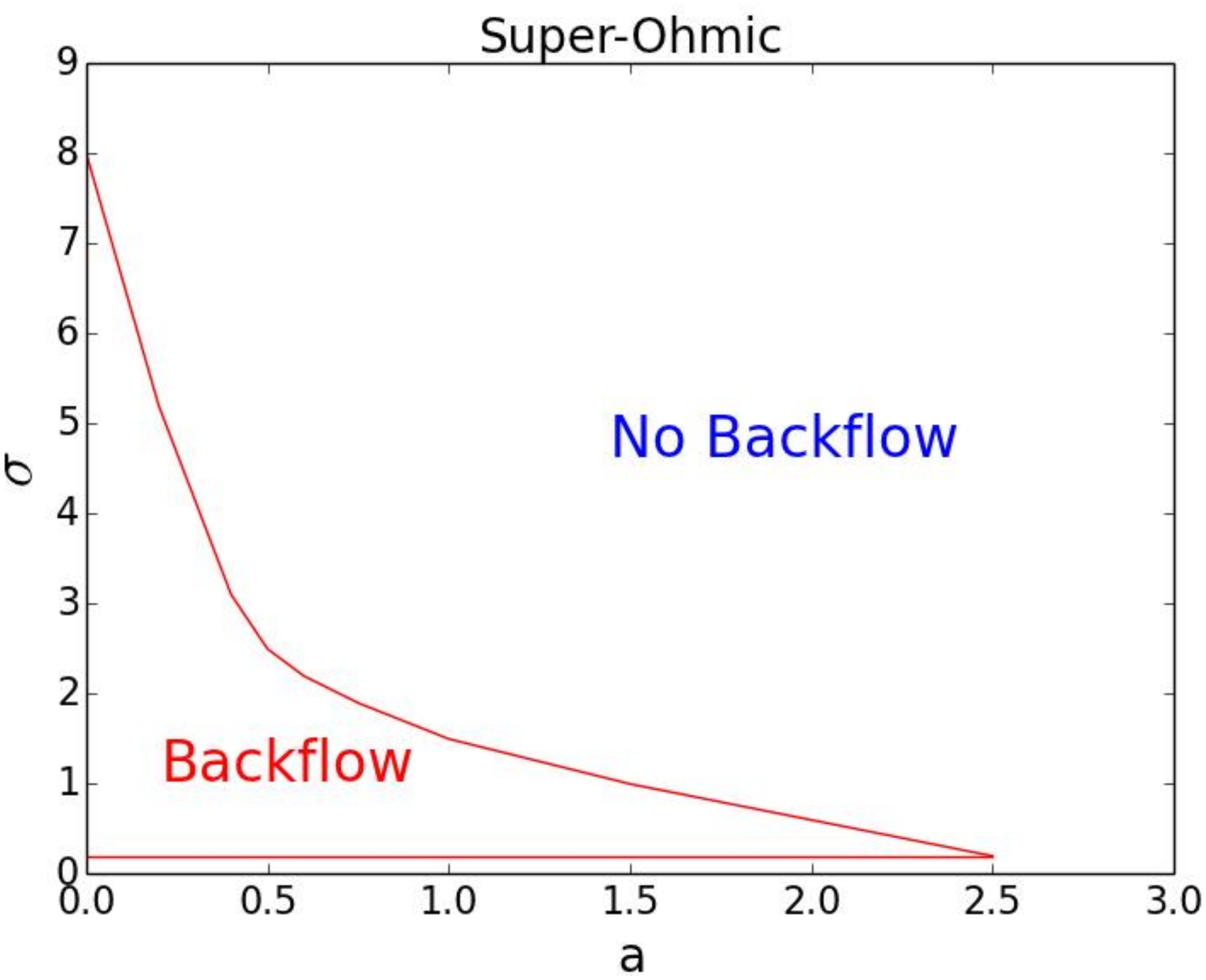}
\caption{The regime of the information backflow in the $M$-frame of a constantly accelerating topoloigcal qubit in the super-Ohmic environment as a function of the time-duration scale  $\sigma$ of the switching function and the acceleration $a$.} 
\label{backflow}
\end{figure}

  Based on the intuition and the studies in the previous sections, the coupling duration $\sigma$ (defined in \eq{sigma-duration} for the switching function) and the size of the acceleration $a$ affect the degrees of thermalization in the most direct and relevant way. We thus plot the parameter space of information backflow by tuning $\sigma$ and $a$. The result is shown  in Fig. \ref{backflow} and it shows that the information backflow is confined in the small $\sigma$ and $a$ regimes.  This justifies our intuition about the relation between thermalization and information backflow.

\subsubsection{Modulation of switching function}

    Naively, the information backflow is a time modulation of the decoherence patterns. We may then wonder if the time modulation of the switching function or acceleration will invoke the information backflow or not. 
    
\begin{figure}
\includegraphics[width=.5\columnwidth]{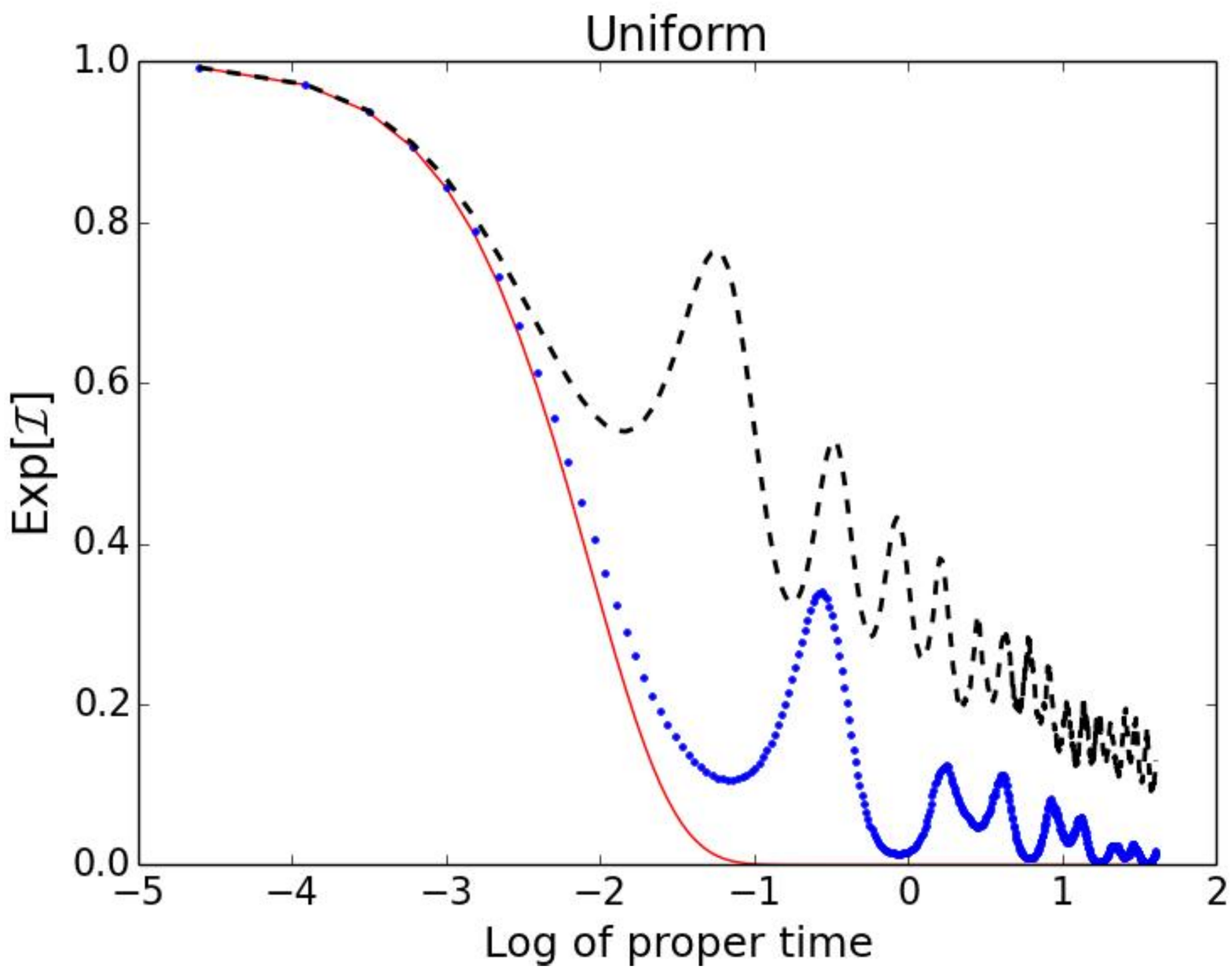}
\includegraphics[width=.5\columnwidth]{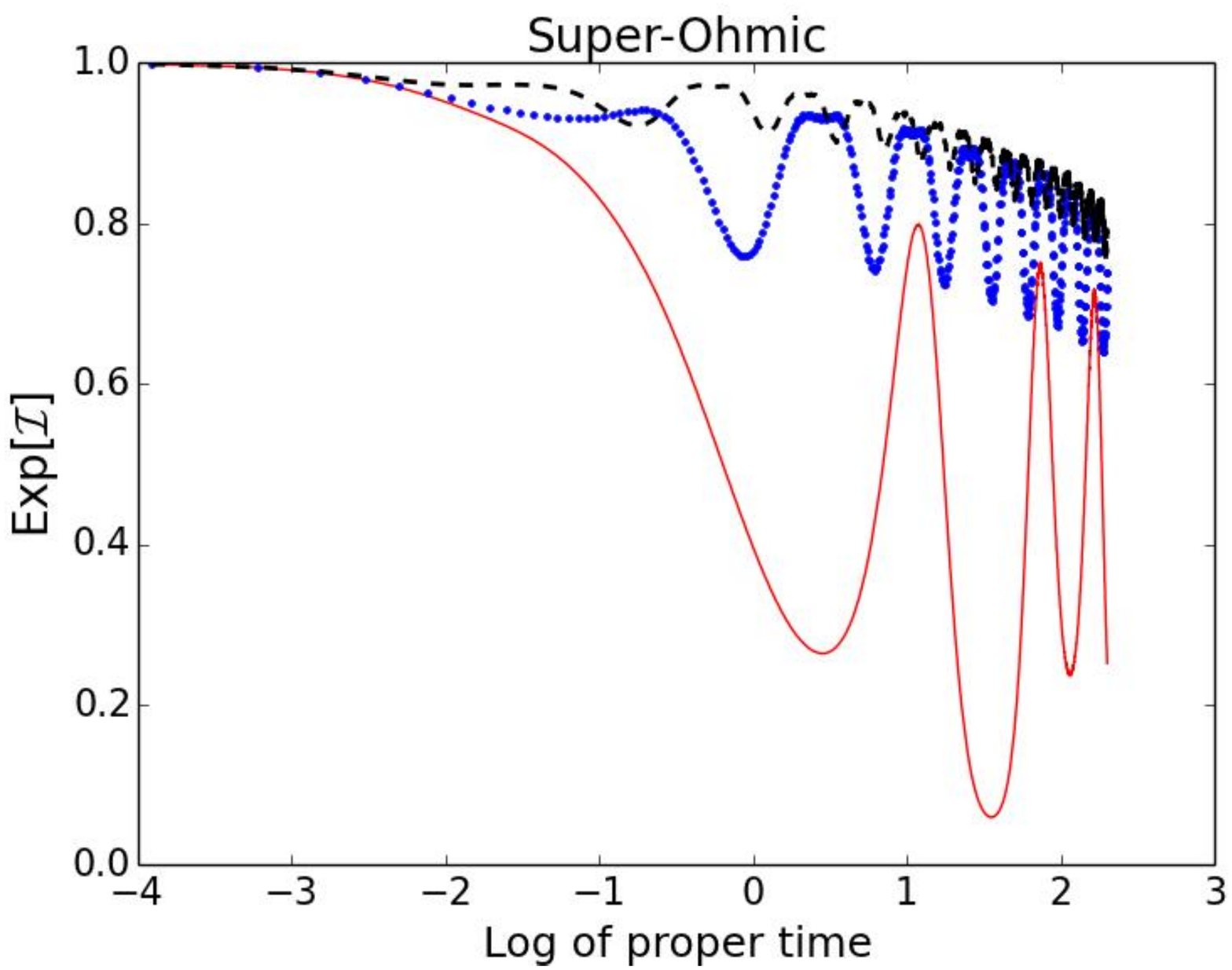}
\caption{Decoherence patterns in the $M$-frame of a constantly accelerating topological qubit  in the environments of uniform spectrum (Left) and of super-Ohmic spectrum (Right) with the frequency modulation of the switching function:  the modulation frequencies are $\omega_M=1$ (solid red), $\omega_M=5$ (dotted blue) and $\omega_M=10$ (dashed black) with the acceleration $a=5$. } 
\label{mod}
\end{figure}

     We first consider the time modulation of the switching function. Instead of using the Gaussian form of the switching function as given in \eq{sigma-duration}, we consider the following form:
\be
\label{swf}
\lambda(\tau)=\cos\omega_M\tau\;.
\ee
The results of the decoherence patterns by varying $\omega_M$ (in some sense, it can be called  ``frequency modulation") are given in Fig. \ref{mod} for the environments of both uniform spectrum and super-Ohmic one.  We find that the decoherence patterns become highly modulated as the modulation frequency $\omega_M$ increases. This phenomenon occurs for both the cases of uniform spectrum and the super-Ohmic one, especially for the former it shows no information backflow if $\omega_M$ is not large enough. 

\subsubsection{Modulation of acceleration}

    We now turn to the effect on the information backflow by by tuning the modulation of acceleration. This was done by considering the general linear motions for which the acceleration is time-varying. IN such cases, the worldline of the trajectory for the Majorana zero mode is given by 
\be
t(\tau)=\int_0^\tau \cosh\big[\int_0^{\tau'}a(\tau'')d\tau''\big]d\tau',\qquad x(\tau)=\int_0^\tau \sinh\big[\int_0^{\tau'}a(\tau'')d\tau''\big]d\tau'
\ee
where $a(\tau)$ is the time-varying acceleration. When $a(\tau)$ is constant, the above trajectory reduces to \eq{constant-a-wl}.

  In the following, we will consider two different forms of $a(\tau)$ and check the corresponding decoherence patterns (with the switching function set to unity).  One is the rectangular function given by 
\be\label{AM-a}
a(\tau) = \left\{\begin{array}{ll}
C, & \mbox{if $\tau_1<\tau<\tau_2$}, \\ 
- C, & \mbox{if $\tau_2<\tau<2\tau_2-\tau_1$}, \\ 
0, & \mbox{otherwise.}
\end{array} \right.
\ee
where $C$ is a constant factor determining the amplitude of the acceleration.  The other is the cosine function given by  
\be\label{FM-a}
a(\tau)=a\cos\omega_G\tau\;.
\ee
We will then tune $C$ in \eq{AM-a} as the ``amplitude modulation" (AM) of the accelerating motion, and tune $\omega_G$ in \eq{FM-a} as the ``frequency modulation" (FM).

\begin{figure}
\includegraphics[width=.5 \columnwidth]{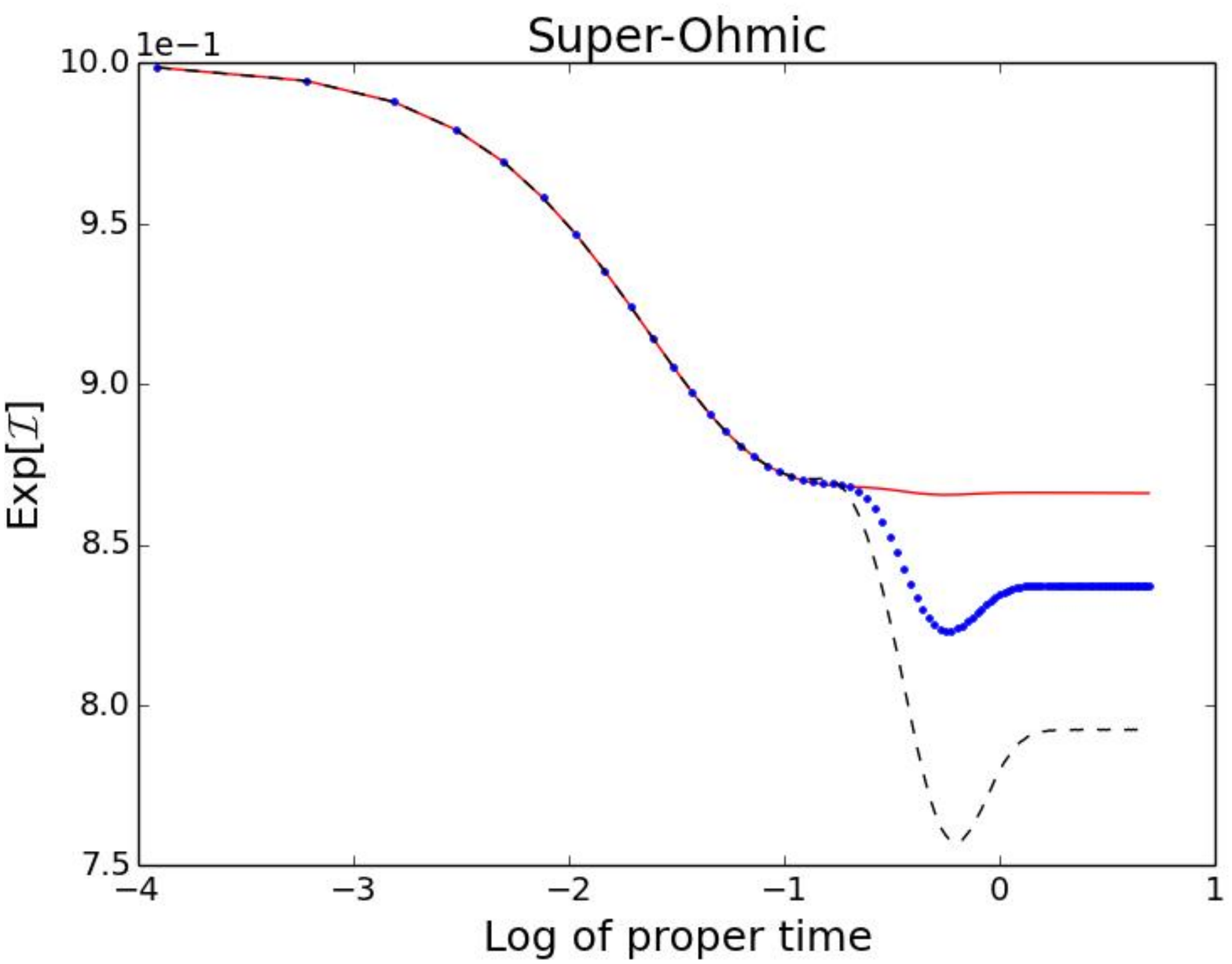}
\includegraphics[width=.5 \columnwidth]{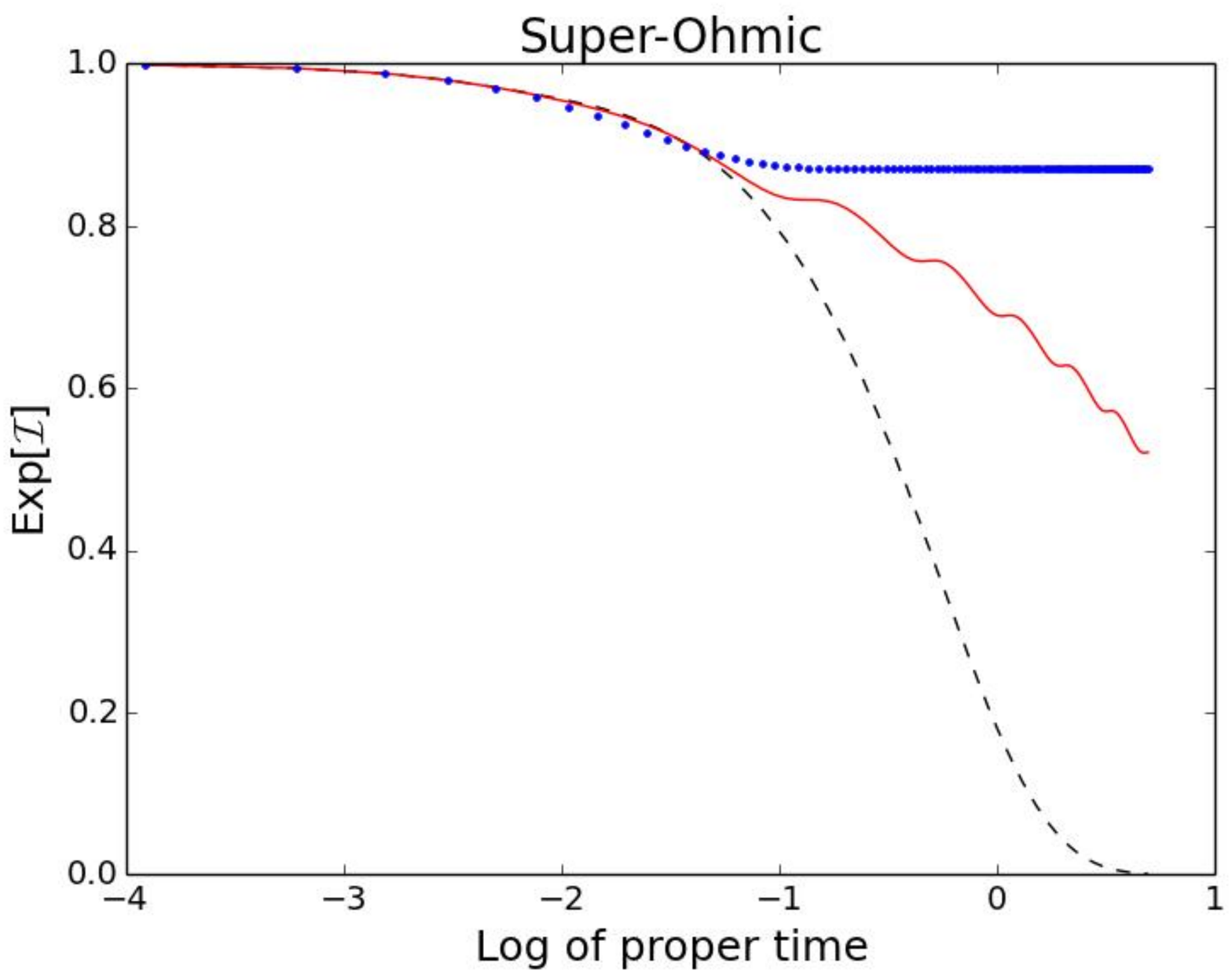}
\caption{Decoherence patterns in the $M$-frame of an accelerating topological qubit in the super-Ohmic environment with the amplitude modulation (AM) and frequency modulation (FM) of the acceleration. Left: AM with $C=1$ (solid red) , $C=5$ (dotted blue) and $C=10$ (dashed black) as $\tau_1=0.3$ and $\tau_2=0.5$. Right:  FM with $\omega_G=10$ (solid red) , $\omega_G=50$ (dotted blue) and $\omega_G=1$ (dashed black) as $a=10$.} 
\label{generic a-lin}
\end{figure}

   The corresponding decoherence patterns of an accelerating topological qubit in the $M$-frame in the super-Ohmic environment for the rectangular (AM) and cosine (FM) modulations of the acceleration are shown in the left and right panels of Fig. \ref{generic a-lin}, respectively.  We see that the under AM the decoherence patterns show the information backflow for the larger amplitude $C$. On the other hand, under FM, it shows no information backflow by tuning the frequency $\omega_G$. However, for the some middle frequency, it shows some ripples in the decoherence pattern but it is still monotonically decaying.  In contrast to the frequency modulation of the switching function, the modulation of the acceleration shows far more mild behaviors of the invoking information backflow.

\subsection{Frame dependence for incoherent motions due to nonlocality}

   For simplicity, so far we have assumed both Majorana modes move with the same worldline. This however does not exploit the nonlocal feature of the topological qubit. To demonstrate the effect of the nonlocality, here we will consider the decoherence patterns of two situations from the comoving frame of one of the Majorana modes, say $\gamma_1$.  This frame is described by the Rindler coordinate $(\tau,\xi)$ relating to the Minkowski ones by \cite{Mukhanov:2007zz}
\be
\label{rindler}
t=(\frac{1}{a}+\xi) \sinh a \tau,\qquad x=(\frac{1}{a}+ \xi) \cosh a\tau - \frac{1}{a}\;.
\ee
Note that the Majorana mode $\gamma_1$ moving with the acceleration $a$ is located at $\xi=0$ so that \eq{rindler} reduces to the worldline \eq{constant-a-wl}. Moreover, to ensure the Rindler and Minkowski times flow along the same direction, it requires $\xi > -{1\over a}$. Also, we will only consider the super-Ohmic environment (with $q=1$) for our demonstrations as the acceleration causes more dramatic change of the decoherence patterns than in the other types of environments.

\subsubsection{Separation by a distance initially}

  The first situation is to consider that the two Majorana modes are separated initially by a distance $L$ but move with the same acceleration, i.e., the worldline of the Majorana mode $\gamma_2$ is given by
\be
\label{rindler2}
t=\frac{1}{a}\sinh a\tau_2,\qquad  x=L+\frac{1}{a}(\cosh a\tau_2-1).
\ee
From \eq{rindler} and \eq{rindler2}, we can then relate the Rindler time $\tau$ of $\gamma_1$ to  $\tau_2$ by the following relation 
\be
\label{rindler4}
\tau_2=\frac{1}{a}\log\frac{a L+\sqrt{a^2 L^2+\coth^2a\tau-1}}{\coth a\tau-1}\;.
\ee

\begin{figure}
\includegraphics[width=.5 \columnwidth]{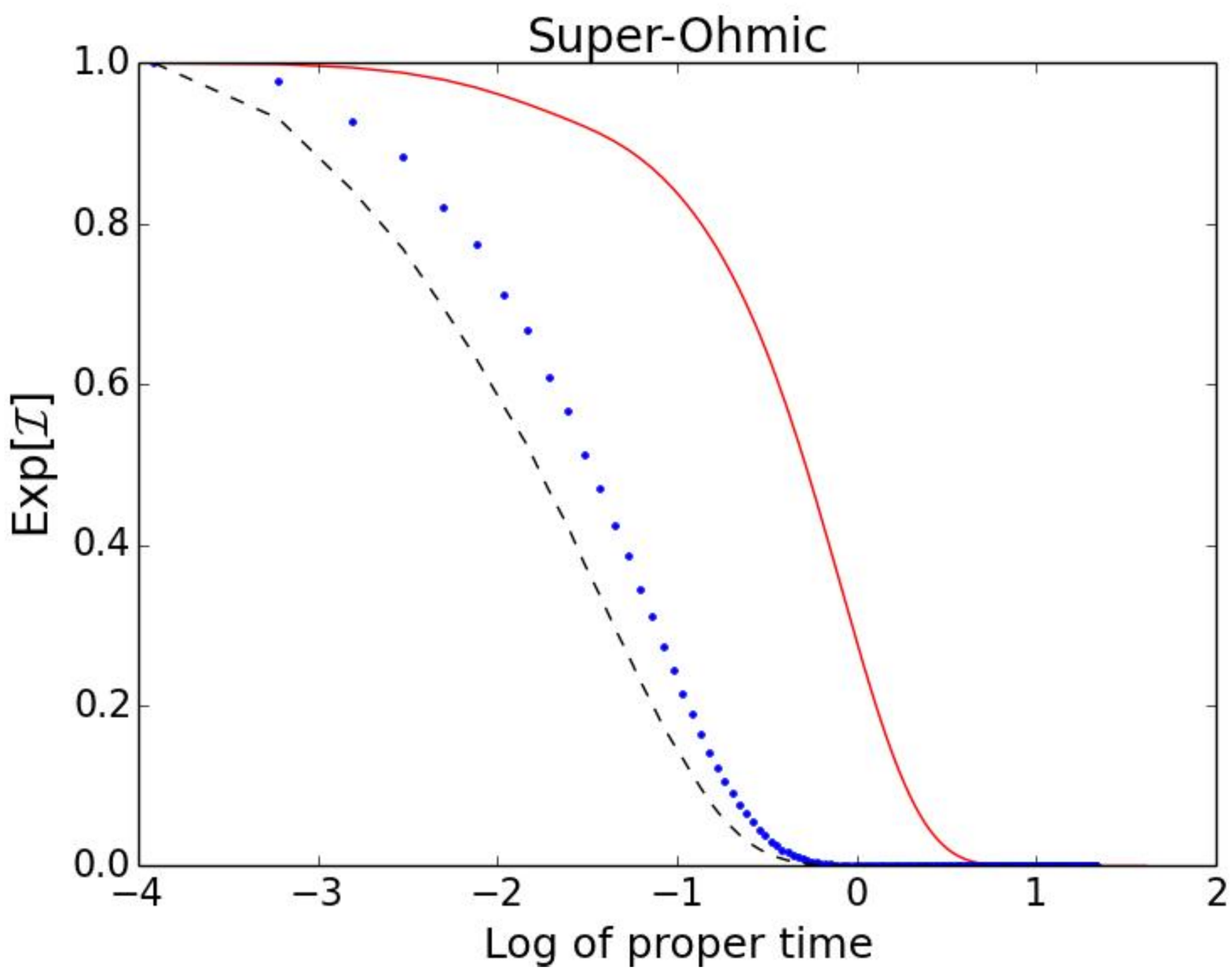}
\includegraphics[width=.5 \columnwidth]{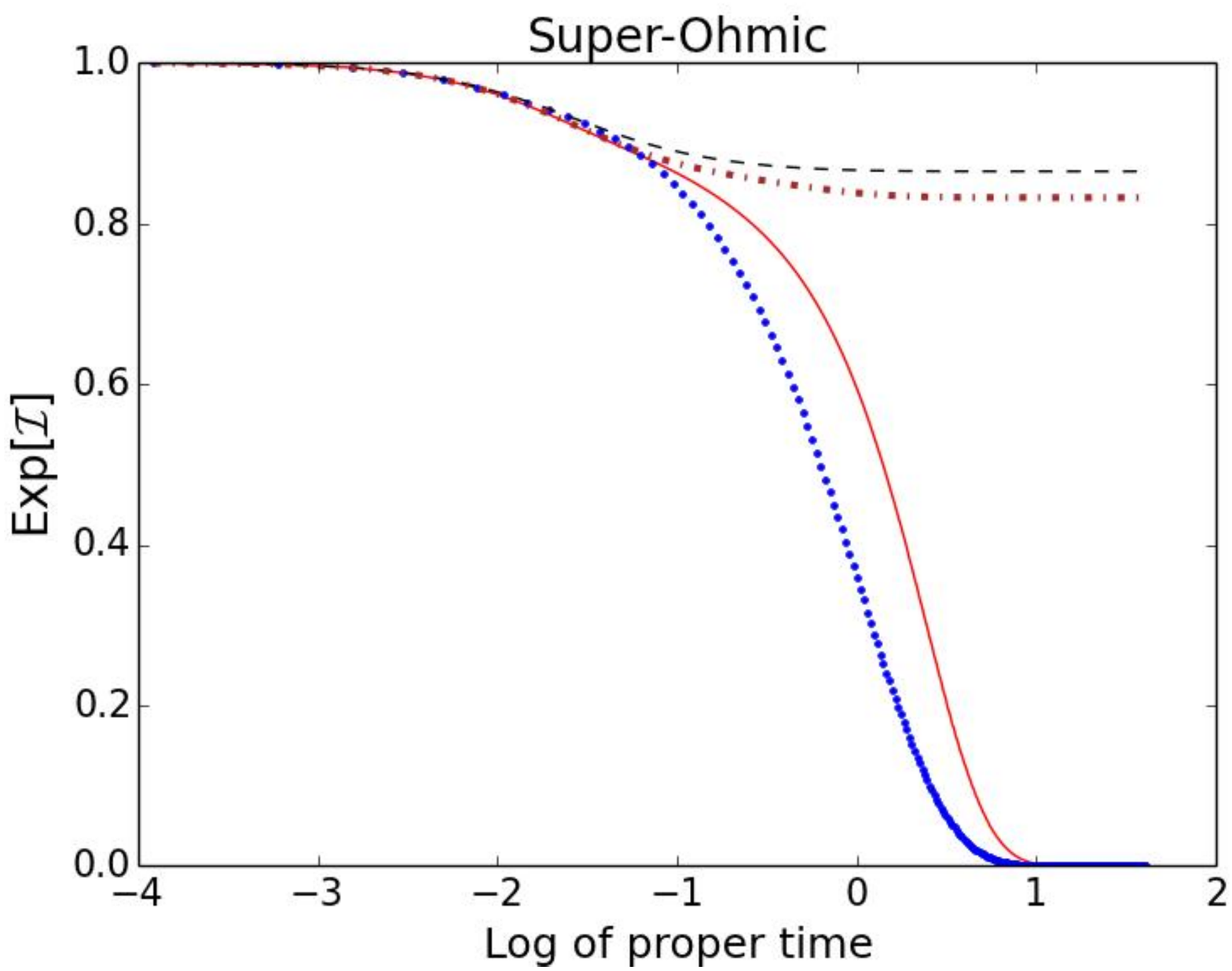}
\caption{Decoherence pattern in the super-Ohmic environment associated with the Majorana mode $\gamma_2$ moving with the acceleration $a_2$ as seen from the observer in the comoving frame of the Majorana mode $\gamma_1$ moving with the acceleration $a$. There is an initial separation $L$ between $\gamma_1$ and $\gamma_2$. Left: $L=0$ (solid red) , $L=1$ (dotted blue) and $L=5$ (dashed black) with $a=a_2=5$. Right:  $a_2=5$ (dotted blue), $a_2=2$ (solid red), $a_2=1$ (dot-dashed brown) and $a_2=-1$ (dashed black) with $a=2$ and $L=0$.} 
\label{diffLA}
\end{figure}

  Using \eq{rindler4} we can evaluate the ``influence functional" $\mathcal{I}_2$ which dictates the decoherence pattern associated with $\gamma_2$ as seen from the observer in the comoving frame of $\gamma_1$. The results are shown in the left panel of Fig. \ref{diffLA}. It indicates that as the initial separation $L$ becomes larger, the rate of decoherence dictated by $\mathcal{I}_2$ is quicker.

\subsubsection{Majorana modes moving with different accelerations}    

  The second situation is to consider that the two Majorana modes moves with different accelerations, say $a$ for $\gamma_1$ and $a_2$ for $\gamma_2$, i.e.,
\be
\label{rindler3}
t=\frac{1}{a_2}\sinh a_2 \tau_2,\qquad  x=\frac{1}{a_2}(\cosh a_2 \tau_2-1).
\ee
As before, one can obtain the relation between their Rindler times: 
\be
\tau_2=\frac{1}{a_2}\log \frac{\frac{a_2}{a} - 1 + \sqrt{(\frac{a_2}{a}-1)^2+ \coth^2a\tau-1}}{\coth a\tau-1}\;.
\ee
It is easy to see that $\tau_2=\tau$ if $a_2=a$. We can then use this relation to evaluate the ``influence functional" $\mathcal{I}_2$ associated with $\gamma_2$ as seen by $\gamma_2$. \footnote{A caution is that we need to ensure the causal condition ${\partial \tau_2 \over \partial \tau} > 0$ (i.e., $\xi > -{1\over a}$) when evaluating $\mathcal{I}_2$. The explicit form of this condition is given in Appendix \ref{app D}. }

 The results for the corresponding decoherence pattern are shown in the right panel of Fig.\ref{diffLA}.  It implies that rate of decoherence dictated by $\mathcal{I}_2$ is quicker than the one dictated by $\mathcal{I}_1$ if $a_2 > a$. Otherwise, it is slower and even stop to decohere if $a_2< a$.  This indicates that the topological qubit in the super-Ohmic environment may not look thermalized in the comoving frame of one of Majorana modes even both Majorana modes are accelerated. This is a novel feature of the topological qubit.

\section{Reduced dynamics of topological qubit in circular motion} \label{sec V}

  In this section we will consider the topological qubit in the circular motion depicted in the right panel of Fig. \ref{cartoon} where the two constituent Majorana modes sit and move diametrically.  This is a quite different setup from the circular motion of the usual UDW detector where the qubit is treated as a whole. 
  
  In contrast to the linearly accelerating UDW detector for which the the transition probabilities derived from the Minkowski and Unruh vacua are the same, this is however not the case for the UDW detector in the circular motion. Namely, the Unruh-like effect (with nonzero transition probability) happens only for the Minkowski vaccum, not for Rindler one \cite{Bell:1982qr,Bell:1986ir,circularUnruh,Crispino:2007eb}.   Though this issue is interesting, we will not checked it for the topological qubit in this work, and may explore it later. Instead, we will continue to work in the Minkowski vacuum so that we can demonstrate some of the features found in the linear motion cases are generic as also found in the circular motions.   
  
    For simplicity, we will consider the circular motion in $(2+1)$D Minkowski spacetime. \footnote{We check some preliminary cases in $(3+1)$D and find that the results are qualitatively the same as in the $(2+1)$D. } The worldline of the Majorana zero mode $\gamma_i$ in a circular motion of radius $r_0$ and with with a generic angular velocity $\Omega(\tau)$ is given by 
\be\label{circular-wl}
t(\tau_i)=\int_0^{\tau_i}\gamma(\tau') d\tau',\quad x(\tau_i)=r_0\cos \Theta_i(\tau_i),\quad y(\tau_i)=r_0\sin \Theta_i(\tau_i)\;,
\ee
with
\be
\Theta_i(\tau_i):=\int_0^{\tau_i}\gamma(\tau') \Omega(\tau')d\tau'+\pi\delta_{i,2}, \qquad \gamma(\tau):=\frac{1}{\sqrt{1-r_0^2 \Omega(\tau)^2}}\;.
\ee     
The factor of $\pi \delta_{i,2}$ reminds the fact that the two Majorana modes sit diametrically to each other.  Note that the angular velocity should be constrained by special relativity, i.e., $r_0\Omega(\tau) \le 1$.   

   For simplicity, in the following we will simply set $r_0=1$. We will also work in the super-Ohmic environment (with $q=1$) to demonstrate the change of the decoherence patterns due to the circular motion.    

\subsection{``Overtaking" phenomenon without frame issue: Constant angular velocity}

\begin{figure} 
\includegraphics[width=1.\columnwidth, height=.5 \columnwidth]{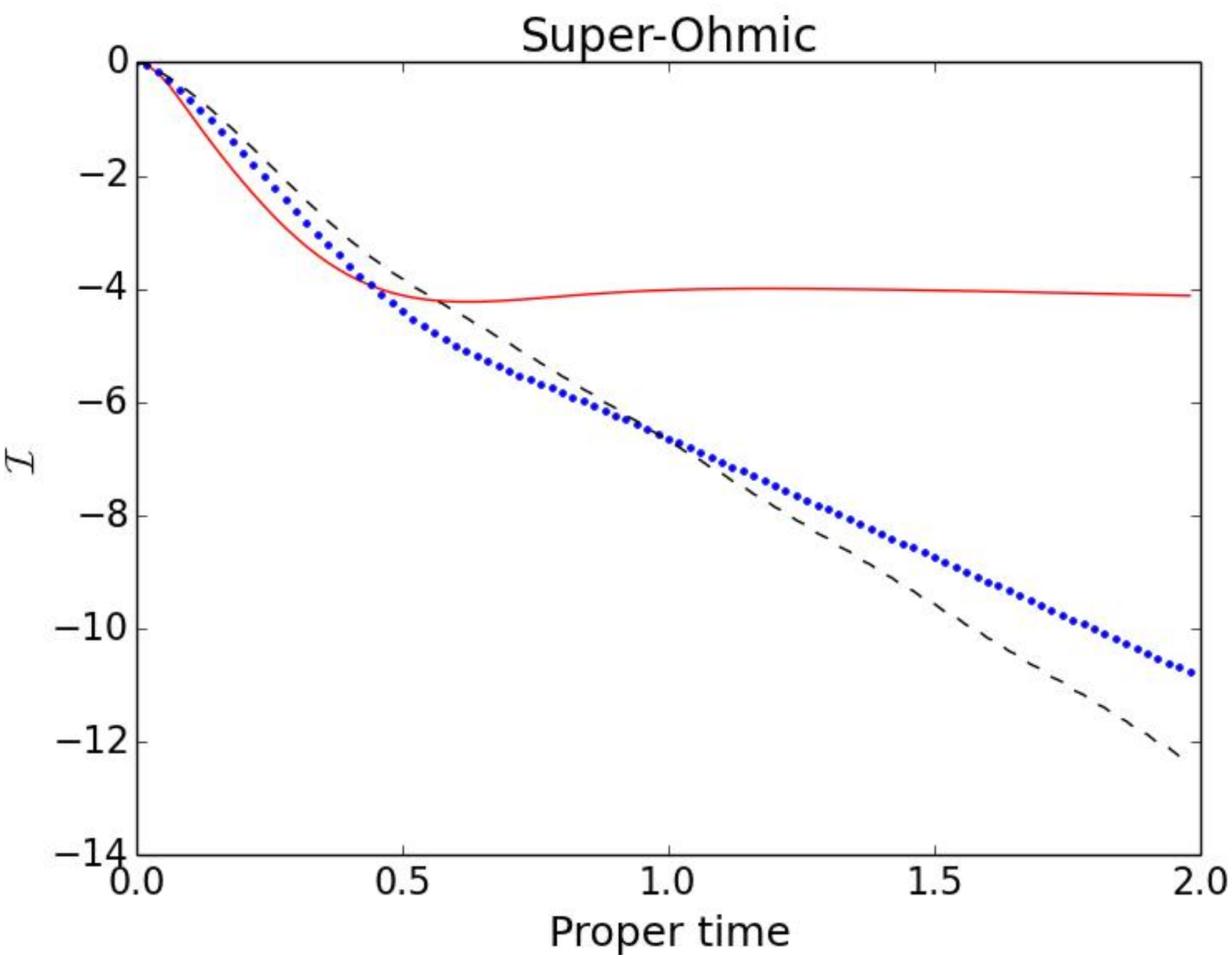}
\caption{``Influence functionals" in the $M$-frame of a circularly moving topological qubit in the super-Ohmic environment with constant angular velocities: $\Omega=0.7$ (solid red), $\Omega=0.9$ (dotted blue) and $\Omega=0.95$ (dashed black). Note that the decoherence patterns show the ``overtaking" phenomenon. } 
\label{const-cir}
\end{figure} 

   We first consider the decoherence patterns for the cases with constant angular velocity and trivial switching function, i.e., $\Omega(\tau)=\Omega, \lambda(\tau)=1$. Then, using the circular worldline \eq{circular-wl} we can evaluate the ``influence functional" \eq{I-M-0} in the $M$-frame. The results are shown in Fig. \ref{const-cir}.  It indicates that there are also the ``overtaking" phenomena as in the linear cases. The subtle difference is that the ``overtaking" does not almost occur at the same time as in the linear motions.

   In the above, we did not specify sub-index for the ``influence functional", and we may wonder if there is a frame-dependence issue as discussed for the linear case due to the incoherent motion and non-locality. In fact, the circular motion in our setup implies that both Majorana modes move coherently. Thus, we shall expect both ``influence functionals" are the same, i.e., $\mathcal{I}_1=\mathcal{I}_2$ as seen in the comoving frame of one of the Majorana zero modes, say $\gamma_1$. Indeed, this is the case.  To see this, we need to introduce the Rindler-like coordinates for the circular motion as given in \cite{Mashhoon:2002xs}, and then relating the Rindler times of two Majorana zero modes. As the procedure is similar to the case for the linear acceleration, we will relegate the details to the Appendix \ref{app D} and just give the result:
\be
\tau_2=\tau+\frac{1}{\gamma \Omega}\left[\cos^{-1}(1-r_0^2\Omega^2)-\pi\right]\;.
\ee
Note that they differ just by a constant shift.  Plugging this relation into \eq{circular-wl} and using the result to evaluate the integrand of the ``influence functional" $\mathcal{I}_2$ of \eq{I-M-0}, it is straightforward to verify $\mathcal{I}_1=\mathcal{I}_2$. \footnote{The explicit derivation goes as follows: {\small
\beq
\mathcal{I}_2(\tau) &:=& -2 \int_0^\tau d\tau_1\int_0^\tau d\tau_2\int_0^{2\pi}d\theta \int_{-\infty}^\infty d\omega|\omega| \mathcal{A}(\omega)e^{-i\omega (t_1-t_2)+i\omega\cos\theta( x_1-x_2)+i\omega\sin\theta( y_1-y_2)}\nn \\
&=&-2\int d\tau_1 d\tau_2 d\theta d\omega |\omega| \mathcal{A}(\omega)e^{-i\omega \gamma(\tau_1-\tau_2)+ i\omega\{ \cos\theta\left[\cos(\gamma\Omega\tau_1+C)-\cos(\gamma\Omega\tau_2+C)\right]+\sin\theta \left[\sin(\gamma\Omega\tau_1+C)-\sin(\gamma\Omega\tau_2+C\right] \}}\nn \\
&=&-2 \int d\tau_1 d\tau_2 d\theta d\omega |\omega| \mathcal{A}(\omega)e^{-i\omega \gamma(\tau_1-\tau_2)+i\omega\left[\cos(\gamma\Omega\tau_1+C-\theta)-\cos(\gamma\Omega\tau_2+C-\theta)\right]} \nn \\
&=&-2 \int d\tau_1 d\tau_2 d\theta' d\omega |\omega| \mathcal{A}(\omega)e^{-i\omega \gamma(\tau_1-\tau_2)+i\omega\left[\cos(\gamma\Omega\tau_1-\theta')-\cos(\gamma\Omega\tau_2-\theta')\right]}=\mathcal{I}_1(\tau)\;.
\eeq}
}.

\subsection{Decoherence Impedance and ``Anti-Unruh"}

\begin{figure}
\includegraphics[width=.5\columnwidth]{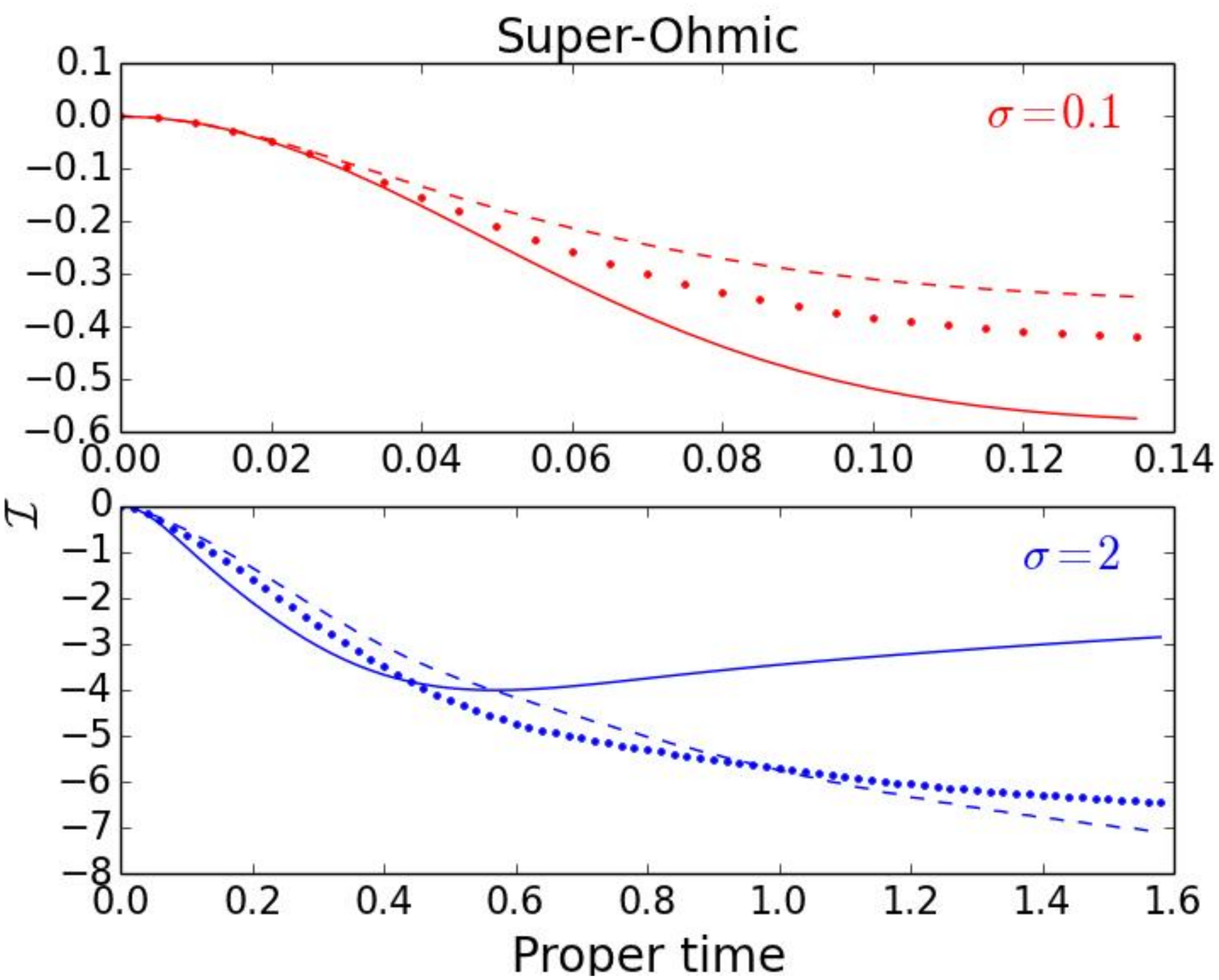}
\includegraphics[width=.5\columnwidth]{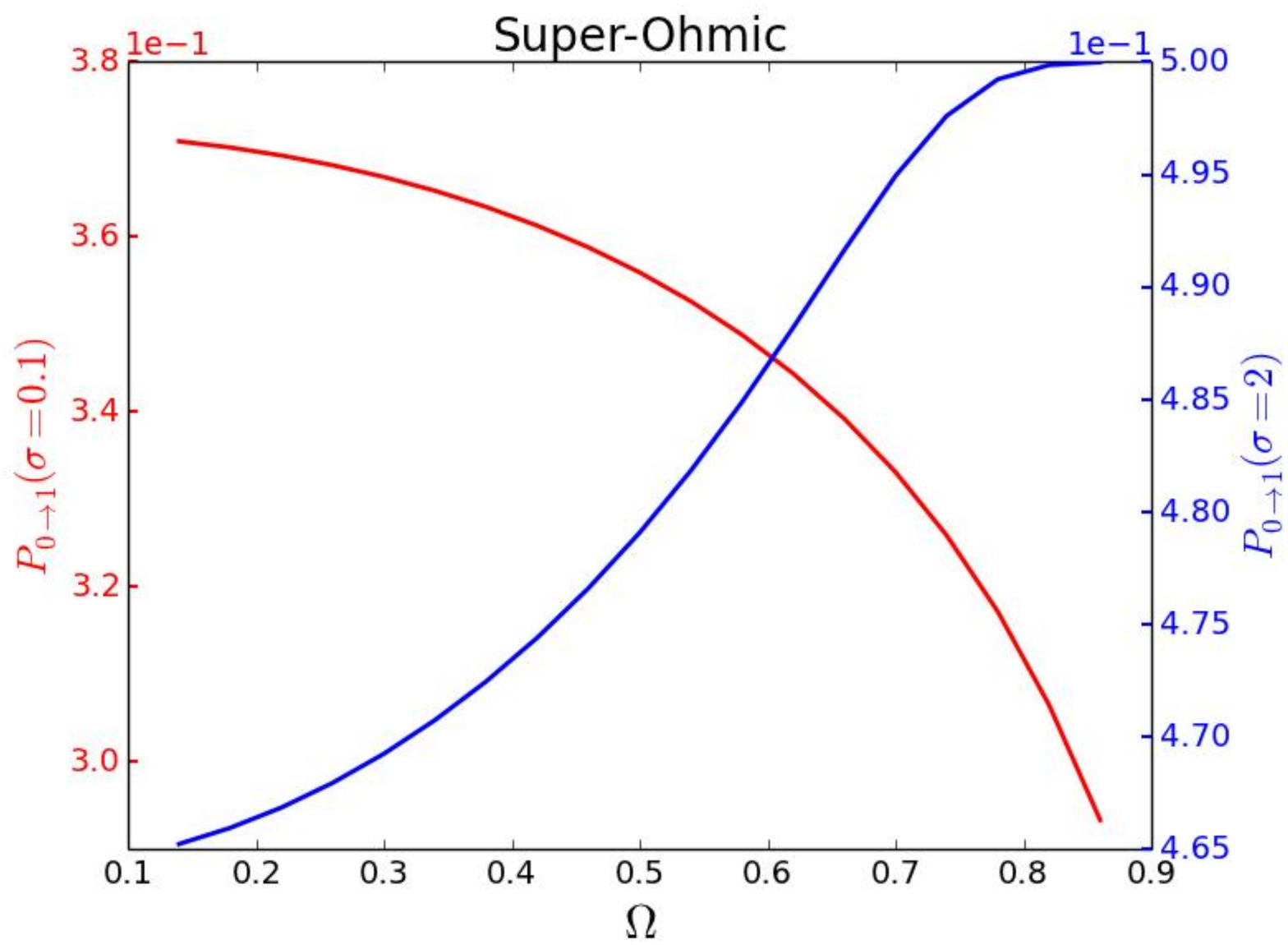}
\caption{ Decoherence patterns and transition probability $P_{0\rightarrow 1}$ in the $M$-frame of a circularly moving MUDW detector of constant angular velocity $\Omega$ in the super-Ohmic environments with    the switching function of time duration scales: $\sigma=0.1$ (red) and $\sigma=2$ (blue). Left : Decoherence patterns for $\Omega=0.7$ (solid), $\Omega=0.9$ (dashed) and $\Omega=0.95$ (dotted). Right:  $P_{0\rightarrow 1}$ versus $\Omega$.  This figure shows that the ``overtaking" and ``anti-Unruh" imply each other.} 
\label{cir-sigma}
\end{figure}

  In order to see if the feature of decoherence impedance and its relation to the ``anti-Unruh" phenomenon found in the linear motion are generic or not, we check here for the circular motion by tuning the the angular velocity for the switching function of finite time-duration, i.e., \eq{sigma-duration}. Again, we find the similar features as found in the linear motion, and the results are shown in Fig.\ref{cir-sigma}. Therefore, we can conclude that these features and phenomena are in some sense quite generic as they appear for both linear and circular motions of the topological qubit.

\subsection{Modulation of switching function}

\begin{figure}
\includegraphics[width=1.\columnwidth, height=.5 \columnwidth]{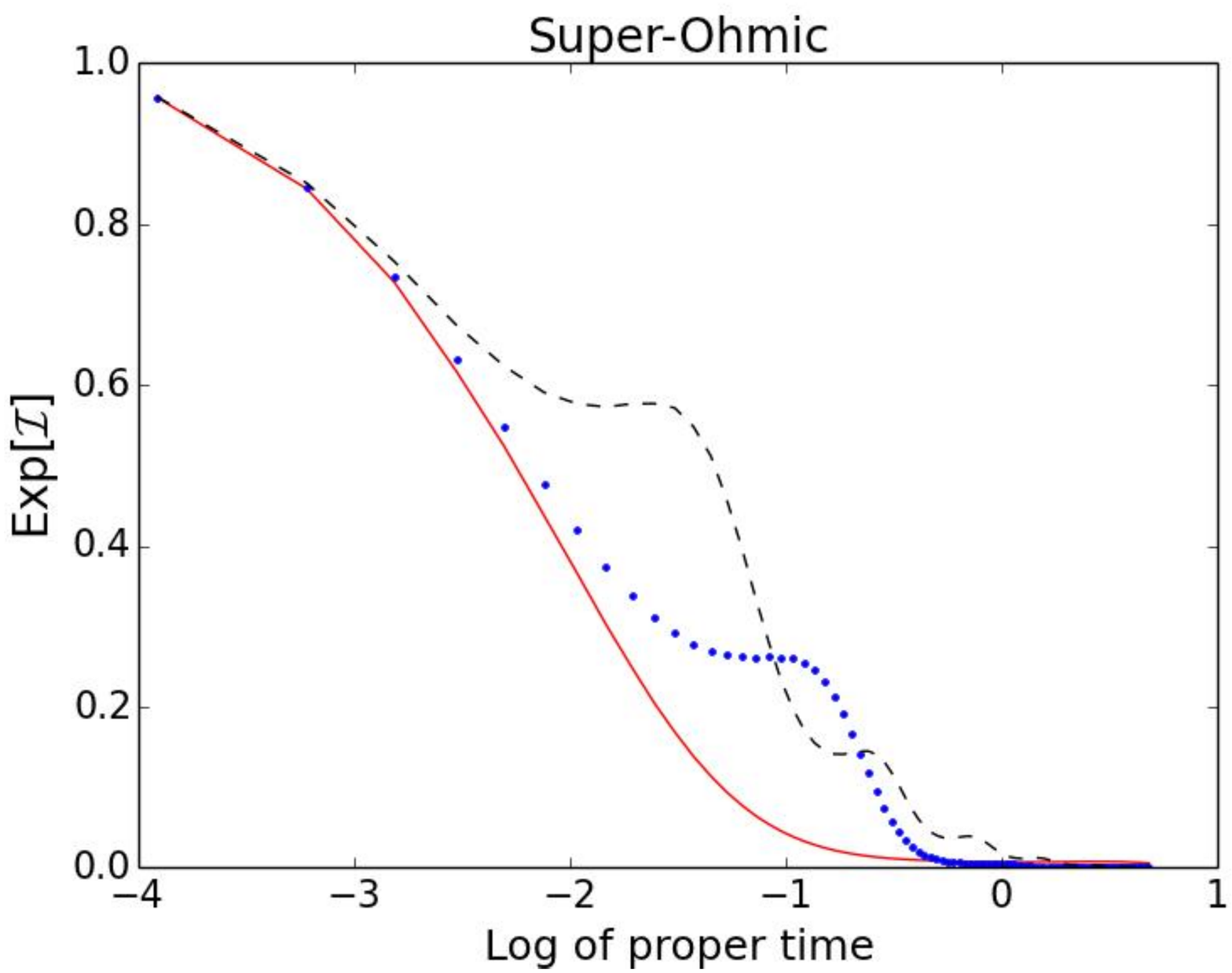}
\caption{Decoherence patterns in the $M$-frame of a circularly moving topological qubit of constant angular velocity $\Omega=0.9$ in the super-Ohmic environment with the frequency modulation of the switching function:  the modulation frequencies are $\omega_M=1$ (solid red), $\omega_M=5$ (dotted blue) and $\omega_M=10$ (dashed black). } 
\label{mod2}
\end{figure}  

   In the linear acceleration case we see that there a very apparent pattern of information backflow as shown in Fig. \ref{mod} when we have the frequency modulation of the switching function by using \eq{swf}.  Here we adopt the same modulation of the switching function in the case of the circular motion.  The results are shown in Fig.\ref{mod2}. Unlike the linear acceleration case, the pattern of the information backflow is barley seen though the decoherence patterns are indeed modulated.   In some sense, it is more similar to the case of the frequency modulation of the acceleration shown by the solid red line in the right panel of Fig. \ref{generic a-lin}.

\subsection{Modulation of angular velocity}

\begin{figure}
\includegraphics[width=.5 \columnwidth]{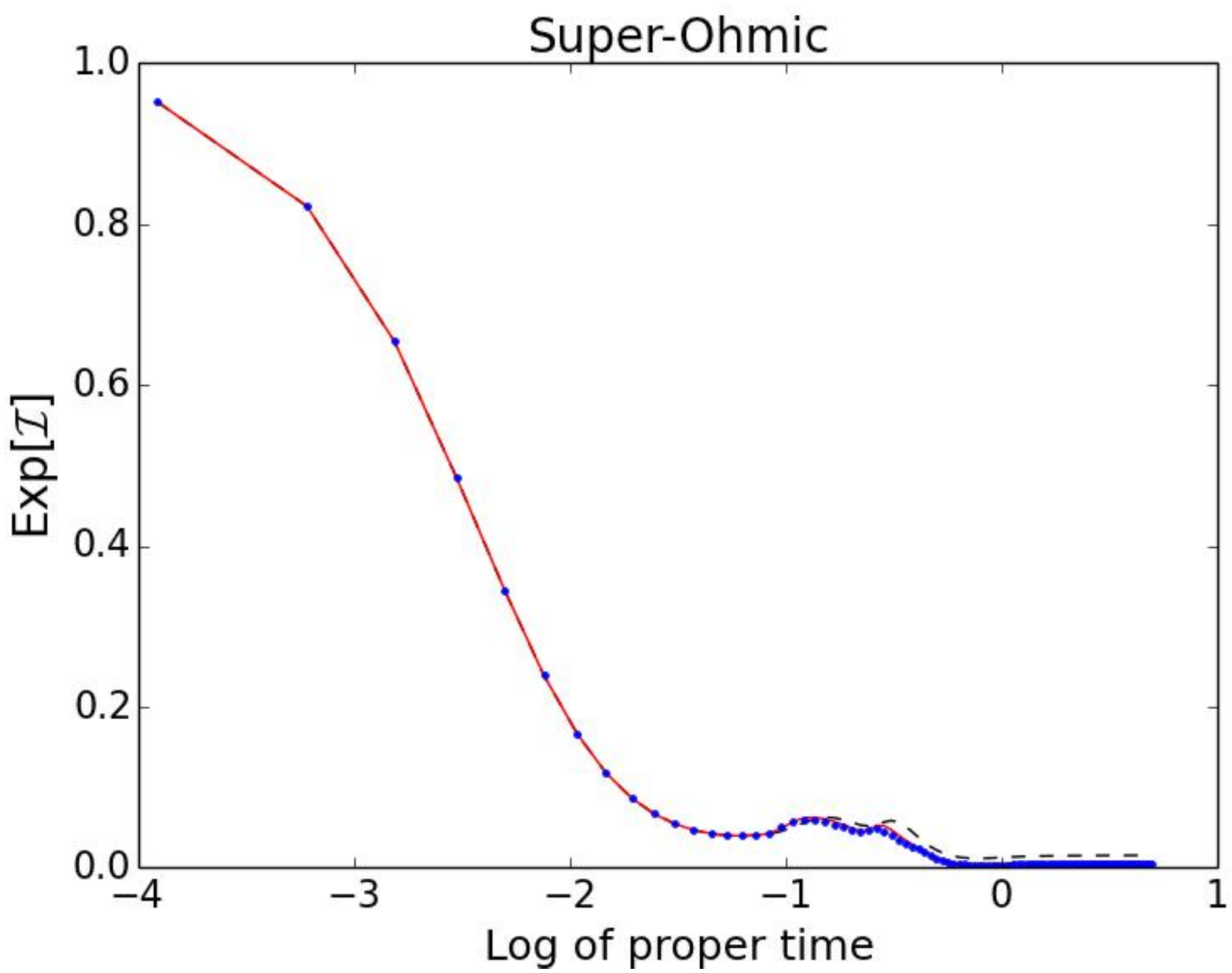}
\includegraphics[width=.5 \columnwidth]{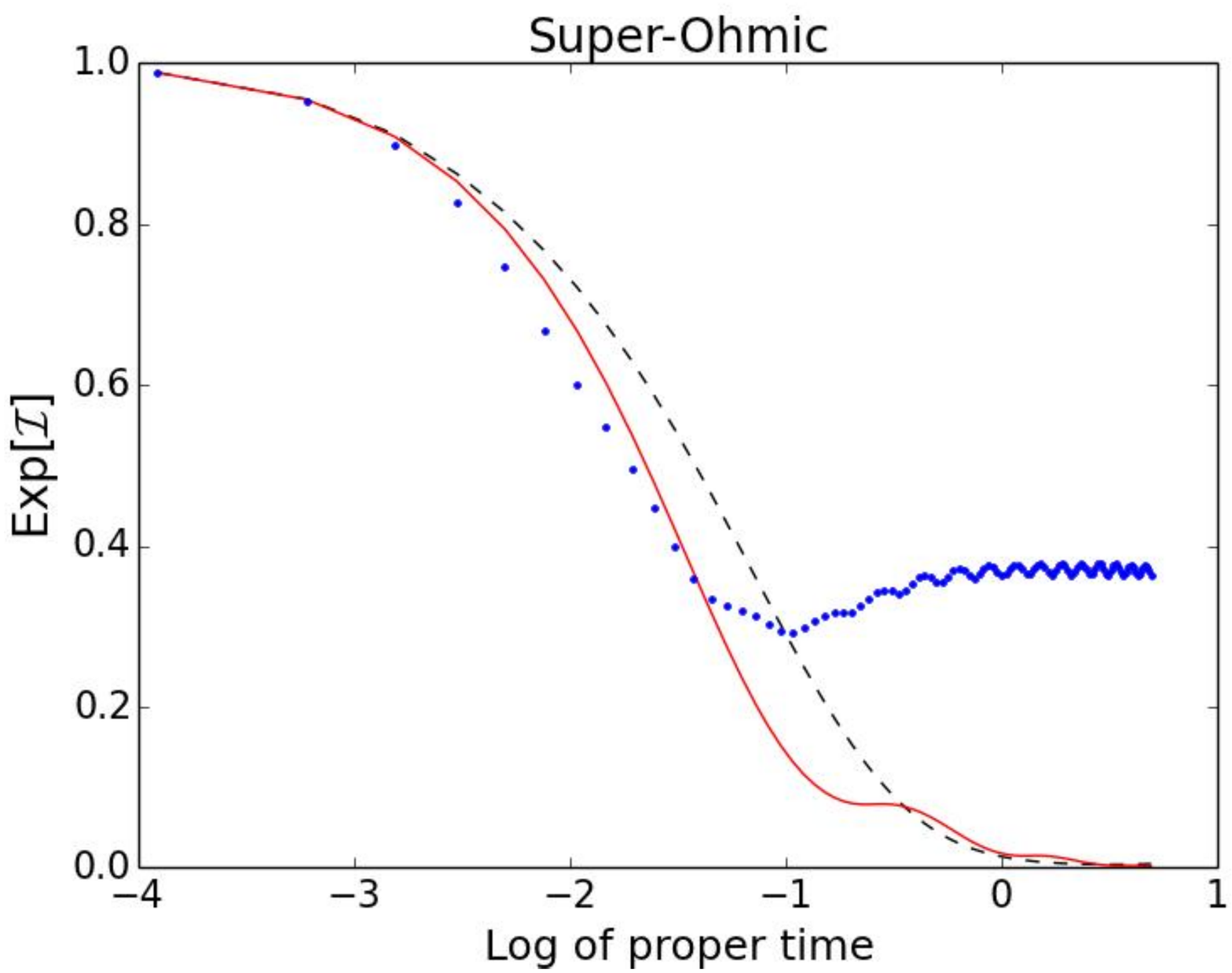}
\caption{Decoherence patterns in the $M$-frame of a circularly moving topological qubit  in the super-Ohmic environment with the amplitude modulation (AM) and frequency modulation (FM) of the angular velocity. Left: AM with $C=0.7$ (solid red) , $C=0.9$ (dotted blue) and $C=0.95$ (dashed black) as  $\tau_1=0.3$, $\tau_2=0.5$. Right:  FM with $\omega_G=1$ (dashed black), $\omega_G=10$ (solid red) and $\omega_G=50$ (dotted blue) as $\Omega=0.95$.} 
\label{generic a-cir}
\end{figure}  

  Finally, we will consider the amplitude and frequency modulations of the angular velocity and see if these modulations will induce the information backflow or not.  In order to compare with the same situations in the linear acceleration cases, we will adopt the same AM and FM functions \eq{AM-a} and \eq{FM-a} for the angular velocity. Note that the $a$ on the RHS of \eq{FM-a} will be replaced by $\Omega$, and we still use $C$ and $\omega_G$ to parametrize AM and FM, respectively.  The results are shown in Fig. \ref{generic a-cir}.  For the AM, we find that there is a minor information backflow. However, it always decoheres completely and the decoherence pattern is not sensitive to $C$ unlike its counterpart of the linear acceleration. For the FM, the results indicates that the higher frequencies induce more apparent information backflow and maintain the robustness against decoherence at the same time. This is also different from its linear counterpart.

\section{Conclusion}\label{secCon}

    In this paper we take the advantage of exact solvable reduced dynamics of the topological qubit and study the decoherence patterns when the topological qubit is subjected to the linear or circular motions. Our results show interesting interplay of relativistic quantum information and the topological ordered states manifested in the Majorana zero modes. 
    
    Though our results are pertinent to topological qubit, we believe the following aspects are generic and should hold even for the usual qubit. They are (1) thermalization due to acceleration; (2) ``anti-Unruh" phenomenon and decoherence impedance found in this paper and their intimate relation as they are both related to the short-time scale non-equilibrium effect; (3) information backflow invoked by the time modulation of coupling constant and accelerations.  We think it deserves more study to clarify the underlying physics for some of the phenomenon and check their generality. 
    
    Besides, by exploiting the nonlocal feature of the topological qubit, we find that some incoherent accelerations of its constituent Majorana zero modes will preserve the coherence. This novel feature may help to develop the robust qubit in the future. Moreover, in this work we only consider the reduced dynamics of single topological qubit. It is interesting to study the multiple topological qubits, especially the evolution of their entanglement.

\acknowledgments
 This work is supported by MoST grant:103-2112-M-003 -001 -MY3  and 103-2811-M-003 -024. We thank Hsi-Sheng Goan, Shih-Yuin Lin, Shin-Tza Wu and Wei-Min Zhang for interesting discussions. We also thank NCTS for partial support. 

\appendix

\section{Evaluation of influence functional by the merging formula of OPE}\label{app A}
In this appendix,  we show how to  express $C_i$ and $S_i$ given in \eq{CiSi}  in term of the "dressed" symmetric Green function $\overline{G}_{i,sym}$ by invoking the following merging formula of OPE \cite{Polchinski}:
\beq
\label{Wick1}
\hat{\cF}&&=e^{\frac{1}{2}\int_0^\tau d\tau_1\int_0^\tau d\tau_2\langle \cO_i(\tau_1) \cO_j(\tau_2)\rangle\frac{\delta^2}{\delta \cO_i(\tau_1)\delta \cO_j(\tau_2)}}:\hat{\cF}:\\
\label{Wick2}
:\hat{\cF}::\hat{\cG}:&&=e^{\int_0^\tau d\tau_1\int_0^\tau d\tau_2\langle \cO_\cF(\tau_1) \cO_\cG(\tau_2)\rangle\frac{\delta^2}{\delta   \cO_\cF(\tau_1)\delta \cO_\cG(\tau_2)}}:\hat{\cF}\hat{\cG}:
\eeq

  Recall $C_i=\langle \mathcal{T}^{\dagger} \cosh {\bf O}_i (\tau) \mathcal{T} \cosh {\bf O}_i(\tau) \rangle$, we can simplify it as follows:
\beq
C_i&&=\frac{1}{4}\langle\mathcal{T}^\dagger(e^{\textbf{O}_i(\tau)}+e^{-\textbf{O}_i(\tau)})\mathcal{T}(e^{\textbf{O}_i(\tau)}+e^{-\textbf{O}_i(\tau)})\rangle \nn \\
&&=\frac{1}{4}e^{\frac{1}{2}\langle \mathcal{T}^\dagger\textbf{O}_i(\tau)\textbf{O}_i(\tau) \rangle+\frac{1}{2}\langle \mathcal{T}\textbf{O}_i(\tau)\textbf{O}_i(\tau) \rangle}\langle (:e^{\tilde{\textbf{O}}_i(\tau)}:+:e^{-\tilde{\textbf{O}}_i(\tau)}:) (:e^{\textbf{O}_i(\tau)}:+:e^{-\textbf{O}_i(\tau)}:)\rangle \nn \\
&&=\frac{1}{4}e^{\frac{1}{2}\langle \mathcal{T}^\dagger\textbf{O}_i(\tau)\textbf{O}_i(\tau) \rangle+\frac{1}{2}\langle \mathcal{T}\textbf{O}_i(\tau)\textbf{O}_i(\tau) \rangle}(2 e^{\langle \tilde{\textbf{O}}_i(\tau)\textbf{O}_i(\tau) \rangle}+2 e^{-\langle \tilde{\textbf{O}}_i(\tau)\textbf{O}_i(\tau) \rangle}) \nn\\
&&=\frac{1}{2}e^{\frac{1}{2}\langle \mathcal{T}^\dagger\textbf{O}_i(\tau)\textbf{O}_i(\tau) \rangle+\frac{1}{2}\langle \mathcal{T}\textbf{O}_i(\tau)\textbf{O}_i(\tau) \rangle}\nn \\\nn
&&\quad ( e^{\frac{1}{2}\langle \tilde{\textbf{O}}_i(\tau)\textbf{O}_i(\tau) \rangle+\frac{1}{2}\langle \textbf{O}_i(\tau)\tilde{\textbf{O}}_i(\tau) \rangle}+ e^{-\frac{1}{2}\langle \tilde{\textbf{O}}_i(\tau)\textbf{O}_i(\tau) \rangle-\frac{1}{2}\langle \textbf{O}_i(\tau)\tilde{\textbf{O}}_i(\tau) \rangle})\\ \label{Ciform}
&&=\frac{1}{2}(e^{2\int^\tau d\tau_1\int^\tau d\tau_2\overline{G}_{i,sym}(\tau_1-\tau_2)}+1)\;.
\eeq
In the above, we apply \eq{Wick1} to arrive the second equality and \eq{Wick2} to obtain the third equality along with the fact that  the expectation value of any normal ordered product is zero. Finally, in the fourth equality we have  symmetrized $\langle \tilde{\textbf{O}}_i(\tau)\textbf{O}_i(\tau) \rangle$ as $\frac{1}{2}(\langle \tilde{\textbf{O}}_i(\tau)\textbf{O}_i(\tau) \rangle+\langle \textbf{O}_i(\tau)\tilde{\textbf{O}}_i(\tau) \rangle)$ to guarantee the symmetry between forward and backward time ordering.  To simplify the notations, we denote the operators on the forward-time contour by the ones without tilde, and the ones with tilde on the backward-time contour.

   Follow the similar steps, we can simplify $S_i=\langle \mathcal{T}^{\dagger} \sinh {\bf O}_i (\tau) \mathcal{T} \sinh {\bf O}_i(\tau) \rangle$ to get the following result:
\beq\label{Siform}
S_i&&=\frac{1}{2}(e^{2\int^\tau d\tau_1\int^\tau d\tau_2\overline{G}_{i,sym}(\tau_1-\tau_2)}-1)\;.
\eeq
 
 Plug the results \eq{Ciform} and \eq{Siform} into \eq{reducedSC} we then arrive the final form of the reduced dynamics given by  \eq{frho-2} and \eq{influence-alpha}.

\section{Spectral density and real-time correlators}\label{app B}
The Schwinger-Keldysh Green functions are the real-time two-point functions on the Keldysh contour. For the fermionic fields, they are defined as follows :
\beq
G^{++}(t,t')&&=-i\langle \textrm{T} \psi(t)\psi^\dagger(t')\rangle\;,\\
G^{--}(t,t')&&=-i\langle \textrm{T}^\dagger \psi(t)\psi^\dagger(t')\rangle\;,\\
G^{+-}(t,t')&&=i\langle\psi^\dagger(t')  \psi(t)\rangle\;,\\
G^{-+}(t,t')&&=-i\langle \psi(t)\psi^\dagger(t')\rangle\;,
\eeq
where $\textrm{T}$ and $\textrm{T}^\dagger$ denote the forward and backward time-ordering for the fermionic fields, respectively. They satisfy
\be
G^{++}+G^{--}=G^{+-}+G^{-+}\;;
\ee
and their relations to the retarded Green function $G_R$ and the symmetric Green function $G_{sym}$ are as follows:
\beq
G_{R}&&=G^{++}-G^{+-}=G^{-+}-G^{--}\;,\\
G_{sym}&&=\frac{i}{2}(G^{++}+G^{--})=\frac{i}{2}(G^{+-}+G^{-+})\;.
\eeq

On the other hand, the Majorana-dressed real-time correlators are 
\beq
\overline{G}^{++}(t,t')&&=-i\langle \mathcal{T} \psi(t)\psi^\dagger(t')\rangle\;,\\
\overline{G}^{--}(t,t')&&=-i\langle \mathcal{T}^\dagger \psi(t)\psi^\dagger(t')\rangle\;,\\
\overline{G}^{+-}(t,t')&&=-i\langle\psi^\dagger(t')  \psi(t)\rangle\;,\\
\overline{G}^{-+}(t,t')&&=-i\langle \psi(t)\psi^\dagger(t')\rangle\;,
\eeq
where $\mathcal{T}$ and $\mathcal{T}^\dagger$ denote the forward and backward time-ordering for bosonic fields, respectively. 
They satisfy
\be
\overline{G}^{++}+\overline{G}^{--}=\overline{G}^{+-}+\overline{G}^{-+}\;;
\ee
and their relations to the Majorana-dressed retarded Green function $\overline{G}_R$ and the Majorana-dressed symmetric Green function $\overline{G}_{sym}$ are as follows:
\beq
\overline{G}_{R}&&=\overline{G}^{++}-\overline{G}^{+-}=\overline{G}^{-+}-\overline{G}^{--}\;,\\
\overline{G}_{sym}&&=\frac{i}{2}(\overline{G}^{++}+\overline{G}^{--})=\frac{i}{2}(\overline{G}^{+-}+\overline{G}^{-+})\;.
\eeq

In frequency domain, the spectral density $\mathcal{A}(\omega)$ is related to $\overline{G}^{+-}(\omega)$ and $\overline{G}^{-+}(\omega)$ by the following relations:
\beq
\overline{G}^{+-}(\omega)&&=-i\mathcal{A}(\omega)\Theta(-\omega)\;.\\
\overline{G}^{-+}(\omega)&&=-i\mathcal{A}(\omega)\Theta(\omega)\;.
\eeq
Therefore, the relation between spectral density $\mathcal{A}(\omega)$ and the dressed symmetric Green function $\overline{G}_{sym}$ is
\be
\overline{G}_{sym}(\omega)=\frac{1}{2}\mathcal{A}(\omega)\;.
\ee

\section{Transition probability}\label{app C}\label{app C}

  Here, we explicitly derive \eq{MUDW-P-1} from \eq{firstP}:  
\beq
P^{(1)}_{0\rightarrow 1}&&:=     \sum_{\bf m} \Big| \int^{\infty} d\tau'   \langle 1| \langle {\bf m}| V^{\tau}(\tau')|{\bf 0}\rangle |0\rangle \Big|^2 \nn \\
&&=     \sum_{\bf m} \Big| \int^{\infty} d\tau'   \langle 1| \langle {\bf m}| \sum_i \gamma_i \; \cO^{\tau}_i(\tau')|{\bf 0}\rangle |0\rangle \Big|^2 \nn \\
&&= \sum_{{\bf m},i}\Big| \int^{\infty} d\tau' \langle {\bf m}| \cO^{\tau}_i(\tau')|{\bf 0}\rangle\Big|^2 \nn \\
&&=\sum_{{\bf m},i}\Big\{\Big|\frac{g_{i}({\bf m})}{|{\bf m}|}\int^\infty d\tau' \lambda(\tau')e^{i|{\bf m}| t'-i {\bf m} x'}\Big|^2+\Big|\frac{\tilde{g}_{i}({\bf m})}{|{\bf m}|}\int^\infty d\tau' \lambda(\tau')e^{i|{\bf m}| t'-i {\bf m} x'}\Big|^2\Big\} \nn\\
&&=\lim_{t \rightarrow \infty} -{1\over 2} \Big( \mathcal{I}_1(t)+\mathcal{I}_2(t)\Big)
\eeq

\section{Causal condition in the comoving coordinates}\label{app D}
   In this appendix, we check the causality condition in the comoving coordinates of linear acceleration and circular motion, which ensures the Rindler times of the two Majorana zero modes have the same time direction. 

\subsection{Linear acceleration: position difference}
    In the linear acceleration case, the causality condition is the same as requiring $\xi>-\frac{1}{a}$.  
For the case with two Majorana zero modes separated by a distance $L$ as given by \eq{rindler2}, using the defining equation \eq{rindler} of the Rindler coordinates we can obtain
\be
\xi=\frac{L}{\cosh a\tau}+\frac{1}{a}(\frac{\cosh a\tau_2}{\cosh a\tau}-1)\;.
\ee
As the hyperbolic cosine function is always positive, we find $\xi$ always is bigger than $-\frac{1}{a}$. Thus, there is no constraint for this case.

\subsection{Linear acceleration: acceleration difference}

  For the case of two Majorana modes moving with different accelerations as specified by \eq{rindler3}, the defining equation \eq{rindler} then yields  
\be
\xi=\frac{1}{a_2\cosh a\tau}(\cosh a_2\tau_2-1)+\frac{1}{a}(\frac{1}{\cosh a\tau}-1)\;.
\ee
Thus, the condition $\xi>-\frac{1}{a}$ yields
\be
\frac{1}{a_2}\cosh a_2\tau_2+\frac{1}{a}-\frac{1}{a_2}>0\;.
\ee
This condition is automatically satisfied for $a_2>0$ as $\cosh a_2\tau_2>1$. For $a_2<0$, it yield a constraint as follows:
\be
\tau_2<-\frac{1}{a_2}\cosh^{-1}(1-\frac{a_2}{a})\;.
\ee

\subsection{Circular motion of constant angular velocity}
For circular motion case of constant angular velocity, the  comoving coordinates $(\tau,x_M,y_M)$ (of the Majorana zero mode $\gamma_1$) and the Minkowski coordinates $(t,x,y)$ are related by \cite{Mashhoon:2002xs}
\beq
\label{cir-coor}
t&&=\gamma\tau+r_0\gamma\Omega y_M,\nn\\
x&&=x_M\cos \gamma\Omega\tau-\gamma y_M\sin\gamma\Omega\tau+r_0\cos \gamma\Omega\tau,\nn\\
y&&=x_M\sin\gamma\Omega\tau+\gamma y_M\cos\gamma\Omega\tau+r_0\sin \gamma\Omega\tau\;.
\eeq
The worldline of $\gamma_1$ is obtained by setting $x_M=y_M=0$.

On the other hand, the worldline of the second Majorana zero mode $\gamma_2$ is given by 
\beq
\label{cir-tau2}
t&&=\gamma \tau_2,\nn\\
x&&=r_0\cos(\gamma\Omega\tau_2+\pi),\nn\\
y&&=r_0\sin(\gamma\Omega\tau_2+\pi).
\eeq
From \eq{cir-coor} and \eq{cir-tau2}, we can find the following relation between $\tau_2$ and $\tau$
\be
\label{tau2tau}
\frac{\tau_2-\tau}{\sin\left[\gamma\Omega(\tau_2-\tau)+\pi\right]}=\frac{\Omega r_0^2}{\gamma}\;.
\ee

\end{document}